\documentclass[letterpaper,11pt]{article}

\usepackage[cbgreek]{textgreek}
\usepackage{url}  
\usepackage{bm}

\usepackage{amsthm}
\usepackage{amsfonts}
\usepackage{amsmath}
\usepackage{amssymb}
\usepackage{changes}
\usepackage{graphicx}
\usepackage{epstopdf}
\usepackage{lineno}
\usepackage{setspace} 
\usepackage{verbatim}
\usepackage{hyperref}
\usepackage{authblk}
\usepackage{color}
\usepackage{textcomp}
\usepackage{multirow}
\usepackage{xcolor}
\usepackage{tikz}
\usepackage{pgf}
\usepackage{caption}
\usepackage{subcaption}
\usepackage{helvet}
\usepackage[eulergreek]{sansmath}
\usepackage{afterpage}

\newcommand{\us}{{\fontshape{ui}{\textmugreek}}s}

\newcommand{\tttanneal}{TTT\textsubscript{anneal}}
\newcommand{\ttttotal}{TTT\textsubscript{total}}

\newcommand{\be}{\begin{equation}}
\newcommand{\ee}{\end{equation}}
\newcommand{\bea}{\begin{eqnarray}}
\newcommand{\eea}{\end{eqnarray}}

\usetikzlibrary{plotmarks}

\newlength\figureheight 
\newlength\figurewidth

\DeclareGraphicsExtensions{.pdf,.png,.jpg}

\renewcommand{\epsilon}{\varepsilon}

\title{Benchmarking a quantum annealing processor with the time-to-target metric.}

\author{James~King\thanks{Corresponding author, jking@dwavesys.com}}
\author{Sheir~Yarkoni}
\author{Mayssam~M.~Nevisi}
\author{Jeremy~P.~Hilton}
\author{Catherine~C.~McGeoch}
\affil{D-Wave Systems, Burnaby, BC}

\begin{document}

\maketitle

\begin{abstract}
In the evaluation of quantum annealers, metrics based on ground state success rates have two major drawbacks.  First, evaluation requires computation time for both quantum and classical processors that grows exponentially with problem size.  This makes evaluation itself computationally prohibitive.  Second, results are heavily dependent on the effects of analog noise on the quantum processors, which is an engineering issue that complicates the study of the underlying quantum annealing algorithm.  We introduce a novel ``time-to-target" metric which avoids these two issues by challenging software solvers to match the results obtained by a quantum annealer in a short amount of time.  We evaluate D-Wave's latest quantum annealer, the D-Wave 2X system, on an array of problem classes and find that it performs well on several input classes relative to state of the art software solvers running single-threaded on a CPU.
\end{abstract}

\section{Introduction}

The commercial availability of D-Wave's quantum annealers\footnote{D-Wave,  D-Wave One, D-Wave Two, and D-Wave 2X are trademarks of D-Wave Systems Inc.} in recent years \cite{Harris2010,johnson2011quantum,Boixo2013} has led to many interesting challenges in benchmarking, including the identification of suitable metrics for comparing performance against classical solution methods. Several types of difficulties arise. First, a D-Wave computation is both quantum and analog, with nothing resembling a discrete instruction or basic operation that can be counted, as in classical benchmarking scenarios; with no common denominator for comparison we resort to runtime measurements, which are notoriously transient and unstable. A second set of issues arises from the fact that we compare an algorithm implemented in hardware to algorithms implemented in software; standard guidelines for benchmarking computer platforms, software, and algorithms (\cite{barr1995designing,bartz2010experimental,jain2008art,johnson2002theoretician}) do not consider this mixed scenario.

Another difficulty, addressed in this paper, arises from the fact that the algorithms of interest are heuristics for an NP-hard {\em optimization} problem. Unlike decision problems, where the algorithm returns a solution that can be verified right or wrong, optimization problems allow heuristic solutions that may be better or worse, and that cannot be efficiently verified as optimal (unless P=NP). The performance of a given heuristic on a given input is not described by a single number (time to find the solution), but rather by a curve that describes the trade-off between computation time and solution quality.  Assuming that neither dominates the whole problem space, how do we decide if one curve is better than another? Barr et al.~\cite{barr1995designing}, Bartz-Beielstein and Preuss \cite[\S 2]{bartz2010experimental}, and Johnson \cite{johnson2002theoretician} discuss several ways to define performance metrics in this context. Parekh et al.~\cite{parekh2015benchmarking} consider this question in the context of quantum annealing when discussing the merits of two possible success criteria: producing an optimal solution for 99\% of instances, or always producing a solution within 99\% of optimal.

Various performance metrics have appeared in the literature on empirical evaluation of D-Wave platforms compared to classical software solvers.
The first significant effort at evaluation was the study by McGeoch and Wang \cite{McGeoch2013}, in which D-Wave's Vesuvius 5 and Vesuvius 6 generations of chips were compared against general purpose software solvers using two general input classes and one class that is native to the D-Wave chip structure (known as a Chimera-structured Ising problem). That paper used performance metrics based on best solutions found within fixed time limits.

Subsequently several papers appeared that focus on time to find ground states (optimal solutions). Boixo et al.~\cite{Boixo2013,Boixo2014}, in comparisons to classical algorithms that also return samples of solutions (simulated annealing and simulated quantum annealing), introduced metrics that count samples-to-solution (STS) and measure time-to-solution (TTS), respectively the number of samples and the total time required by a solver to find a ground state with sufficiently high probability; these metrics have emerged as standards used in many subsequent papers (e.g.~\cite{Hen2015,Venturelli2014}). R{\o}nnow et al.~\cite{Roennow2014} describe methods for evaluating ``quantum speedup'' that involve comparisons of scaling curves corresponding to TTS over a range of problem sizes.

Selby \cite{Selby2014} uses a somewhat different approach based on the average time to find a ``presumed ground state'' several times in succession.  Selby's implementation \cite{Selby2013git} of the Hamze-de Freitas-Selby algorithm \cite{Hamze2004,Selby2014} may be used for finding the presumed ground state energies of D-Wave's native Chimera structured Ising Hamiltonians. When applied to one problem class in our test set (RAN7, see Section \ref{sec:problems}) this algorithm takes on the order of 0.1 seconds for the graph of the D-Wave One system (2011), 15 seconds for the graph of the D-Wave Two system (2013), and 17 minutes for the graph of the D-Wave 2X system (2015) \cite{Selby2014}, and still does not guarantee optimality. Running CPLEX (a state-of-the-art commercial optimization solver \cite{cplex2015}) to certify optimality takes several hours or days per input at the largest problem sizes. 

This  illustrates a fundamental difficulty with metrics based on finding ground states: on sufficiently challenging inputs the time required for any solver grows exponentially with problem size\footnote{Using known algorithms or assuming the exponential time hypothesis \cite{impagliazzo1999complexity}.}.    Thus the computation of STS and TTS for large inputs is intractable. 

Some approaches for getting around this difficulty in the context of evaluation are known. For example Hen et al.~\cite{Hen2015} (see also King \cite{King2015}) shows how to construct inputs with ``planted solutions'' so that optimal solutions are known {\em a priori} by construction. This is an effective approach, but in
general it can be difficult to argue that inputs constructed this way are sufficiently challenging, or representative of real-world inputs, to be interesting \cite{johnson2002theoretician,McGeoch2012}. Another idea is to find an easily-computed lower bound on the optimal solution and to use that bound as proxy for optimal (measuring, for example, time to get within a certain percentage of the bound), but nothing suitable appears to be available for the native problems solved by D-Wave quantum annealers.  Saket \cite{Saket2013} developed a PTAS\footnote{Polynomial-time approximation scheme.} for Chimera Ising models, but this is mostly of theoretical interest---to get within a fraction $\varepsilon$ of the ground state energy takes $O(n\cdot 2^{32/\varepsilon})$ time, impractical even to get within 10\%.

A second problem with using ground states in performance metrics is the fact that analog noise in D-Wave's processors perturbs the input Hamiltonian.  This means that the nominal input specified by the user differs slightly from the physical input solved by the processor \cite{Venturelli2014,King2015,Young2013,Pudenz2014,Vinci2014}. This problem is particularly pernicious for finding ground states since ground states and excited states in the nominal Hamiltonian may exchange roles in the physical Hamiltonian, and generally near-ground states far outnumber ground states.

In this study we introduce a new metric that avoids the problem of prohibitive runtimes and makes evaluation far less sensitive to analog noise.  We do this by having the solvers race to a target energy determined by the D-Wave processor's energy distribution.  We call this the ``time-to-target" (TTT) metric.  Our use of the D-Wave processor as a reference solver in computing the TTT metric allows us to circumvent the difficulties of evaluating performance in finding ground states,  and to explore an interesting property that we have observed: very fast convergence to near-optimal solutions.     

The TTT metric identifies low-cost target solutions found by the D-Wave processor within very short time limits (from 15ms to 352ms in this study),  and then asks how much time competing software solvers need to find solution energies of matching or better quality.  We observe that the D-Wave processor performs well both in terms of total computation time (including I/O costs to move data on and off the chip), and pure computation times (omitting I/O costs).  Our results may be summarized as follows.

\begin{itemize}
\item  Considering total time from start to finish (including I/O costs), D-Wave 2X TTT times are 2x to 15x faster than the best competing software (at largest problem sizes), for all but one input class that we tested, in which a solver specific to that input class is faster.

\item Considering pure computation time (omitting I/O costs), D-Wave 2X  TTT times are 8x to 600x faster than competing software times on all input classes we tested. 
\end{itemize} 
 
The next section presents an overview of D-Wave quantum annealing technology and describes the performance models for D-Wave quantum annealers and competing software solvers.   Section \ref{ttt} defines the TTT metric and a related STT (samples to target) metric.  Section \ref{setup} describes our experimental setup,  Section \ref{results} presents experimental results,  and Section \ref{discussion} summarizes our observations and conclusions.   Futher descriptions of the D-Wave 2X,  discussion of our experimental procedures, and results for the full suite of test problems,  may be found in the appendices.  

\paragraph{A note on quantum speedup}
R{\o}nnow et al.~\cite{Roennow2014} searched for ``limited quantum speedup'' by looking for differentiation in scaling trends between the performance of a D-Wave Vesuvius chip and the performance of a classical analog---a theoretical thermal annealer with free parallelism (i.e., simulated annealing with theoretical constant-time sweeps)---in terms of TTS as a function of input size.  We take special care here to emphasize that in this paper we do not address the issue of quantum speedup; rather our goal is to compare runtime performance strictly within the range of problem sizes tested.

\section{Overview} 
\label{sec:overview} 

We start with an overview of  D-Wave design features and introduce notation that will be used throughout.   For details about underlying technologies see Bunyk et al.~\cite{bunyk2014architectural}, Dickson et al.~\cite{Dickson2013}, Harris et al.~\cite{Harris2010},  Johnson et al.~\cite{johnson2011quantum} or Lanting et al.~\cite{Lanting2014}.

\paragraph{Ising Minimization} 
D-Wave quantum annealing processors are designed to find minimum-cost  solutions to the  
 Ising Minimization (IM) problem,  defined  on a graph $G = (V,E)$ as follows.  Given a collection of weights $h = \{ h_i: i \in V \} $ and $J = \{J_{ij}: (i, j) \in E \}$,  
 assign values from $\{ -1, +1 \}$  to  $n$ {\em spin variables}  $s =\{ s_i \}$ 
so as to minimize the {\em energy function}         
\begin{eqnarray}\label{eqn:ising}
 E( s ) &=  & \sum_{i\in V}  h_i  s_i   +  \sum_{(i,j)\in E} J_{ij} s_i  s_j.  
\end{eqnarray}
The spin variables $s$  can be interpreted as magnetic poles in a physical particle system;  in this context,  negative $J_{ij}$ is {\em ferromagnetic} 
and positive $J_{ij}$ is {\em antiferromagnetic},  the optimal solution is called a {\em ground state},  and non-optimal solutions are {\em excited states}.  
IM instances can be trivially transformed to Quadratic Unconstrained Boolean Optimization (QUBO) instances defined on integers $s = \{0,1\}$,
or to Maximum Weighted 2-Satisfiability (MAX W2SAT)  instances defined on booleans {\em  s= \{true, false\}}, all of which are 
NP-hard.   

\paragraph{Chimera topology} 
The native connectivity topology for D-Wave 2X platforms is based on a $C_{12}$ {\em Chimera graph} containing  $1152$ vertices (qubits) and $3360$ edges (couplers).  

A  Chimera graph of size $C_s$ is an $s\times s$ grid of chimera cells, each containing a complete bipartite graph on 8 vertices (a $K_{4,4}$).  Each vertex is connected to its four neighbors inside the cell as well as two neighbors (north/south  or east/west) outside the cell: therefore every vertex has degree 6 excluding boundary vertices.

In this study, as in others, we vary the problem size using square subgraphs of the full graph, from size $C_4$  (128 vertices) up to $C_{12}$ (1152 vertices).  Note that the number of problem variables $n=8s^2$ grows quadratically with Chimera size.  The reason we measure algorithm performance as a function of the Chimera size and not the number of qubits is that problem difficulty tends to scale exponentially with the Chimera size, i.e., with the square root of the number of qubits, since the treewidth of a Chimera graph $C_s$ is linear in $s$ \cite{Boixo2013}.

Because the chip fabrication process leaves some small number (typically fewer than  $5\%$)  of qubits unusable, each processor has a specific  {\em hardware working graph}  $H \subset$ $C_{12}$.  The working graph used in this study has 1097 working qubits out of 1152 (see Appendix \ref{app:hardware} for an image).

\paragraph{Quantum annealing.} 
D-Wave processors solve Ising problems by {\em quantum annealing} (QA),  which belongs to the {\em adiabatic quantum computing} (AQC) paradigm.   The QA algorithm is implemented in hardware using a framework of analog control devices to 
manipulate a collection of qubit states according to a time-dependent Hamiltonian shown below.     
\begin{eqnarray}\label{eqn:aqc}
{\cal H}(t)  &= & A(t)  \cdot {\cal H}_{init}   +B(t) \cdot {\cal H}_{prob}. 
\end{eqnarray} 
This algorithm carries out a gradual transition in
 time $t : 0 \rightarrow t_a$,  from an initial ground state in ${\cal H}_{init}$, to a state described 
by the {\em problem Hamiltonian} ${\cal H}_{prob} = \sum_{i} h_i \sigma^z_i + \sum_{ij} J_{ij} \sigma^z_i \sigma^z_j$.   The problem Hamiltonian   
matches the energy function \eqref{eqn:ising},  so that a ground state for ${\cal H}_{prob}$  is a minimum-cost solution to $E(s)$. 
The AQC model of computation was proposed by Farhi et al.~\cite{Farhi2001} who showed that if the transition is carried out slowly enough the algorithm will find a ground state (i.e.~an optimal solution) with high probability.  

The D-Wave processor studied here has a chip containing 1097 active qubits (quantum bits) and 3045  active couplers made of microscopic loops of niobium connected to a large and complex analog control system via an arrangement of Josephson Junctions.  When cooled to temperatures below 9.3K,  niobium becomes a superconductor and is capable of displaying quantum properties including {superposition},  {entanglement}, and {quantum tunneling}.  
Because of these properties the qubits on the chip behave as a quantum mechanical particle process that carries out a transition from initial state described by ${\cal H}_{init}$ to a problem state described by ${\cal H}_{prob}$ \cite{Boixo2014,Dickson2013,Lanting2014}.

Theoretical guarantees about solution times for quantum algorithms (found in \cite{Farhi2001})  assume that the computation takes place in an ideal closed system,  perfectly isolated from energy interference from ambient surroundings.  The D-Wave 2X chip is housed in a highly shielded chamber and cooled to below 15mK; nevertheless as is the case with any real-world  quantum device,  it must suffer some amount of interference,  which has the general effect of reducing the probability of landing in a ground state.  Thus theoretical guarantees on performance may not apply to these systems.  We consider any  D-Wave processor to be a {\em heuristic} solver,  which requires empirical approaches to performance analysis.

\paragraph{Modeling performance.} 
Given input instance $(h,J)$,  a D-Wave computation involves the following steps.   
\begin{enumerate}
\item {\bf Program.}   Load $(h,J)$ onto the chip; denote the elapsed programming/initialization time $t_i$.   
\item {\bf Anneal.}   Carry out the QA algorithm.  Anneal time $t_a$ can be set by user to some value $20 \text{\us} \leq t_a \leq 20,000\text{\us}$.  

\item {\bf Read.}   Record qubit states to obtain a solution; denote the elapsed readout time $t_r$.
\item {\bf Repeat.}  Iterate  steps 2 through 4,  $R$ times,  to obtain a sample of $R$ solutions.  
\end{enumerate} 
We define {\em sample time}  $R\,t_s$ and  {\em total time} $T$ as follows: 
\begin{eqnarray}
t_s & = &(t_a + t_r) \\ \nonumber
T & = & t_i  +  R\, t_s.  
\end{eqnarray}
%
%
%
%
\paragraph{Performance metrics.} 
As is common with optimization heuristics whether quantum or classical,  performance is characterized by a trade-off between {computation time} and {solution quality}: more time yields better results.     
 
At least two such trade-offs can be identified in the D-Wave performance model:  increasing anneal time $t_a$  and/or increasing the  sample size  $R$ improves the probability of observing low-cost solutions in a sample.   In current technologies,  growing $t_a$ above its $20$\us{} lower bound appears to have very little effect on optimization performance (on the inputs classes we've considered);   in this small study we fix $t_a = 20$\us{} and focus the question of what value of $R$ is needed to achieve a target solution quality.  

The STT metric defined in the next section provides an expected value for $R$ based on analysis of the distribution of 
solution samples returned by the D-Wave 2X processor chip.  The TTT metric calculates expected time (anneal time or total time) to reach a target solution quality that depends on the expectation STT.   

\section{The time-to-target (TTT) metric}
\label{ttt} 

TTT is defined with respect to a parameter $q$, a specific reference solver (in this case the D-Wave 2X), and a specific input as follows.  First, we gather many samples from the reference solver.  Then, based on the energy distribution of these samples, we identify the \emph{target energy} $E_q$ that is found at quantile $q$ of the sample distribution.  For example, in this study we take 50,000 samples from the reference solver for each input; for $q=0.001$, we would sort the 50,000 sample energies from lowest to highest and take the 50th as the target energy.  

We want this energy to be on the lower tail of the energy distribution, so that it is non-trivial to reach, but not so low that it is an extreme outlier.

For each input, target energy $E_q$, and solver, we define the samples-to-target (STT) as the expected number of samples required by the solver to attain an energy at or below the target energy.

\begin{figure}
\begin{minipage}[b]{0.48\linewidth}
\centering
\includegraphics[width=\linewidth]{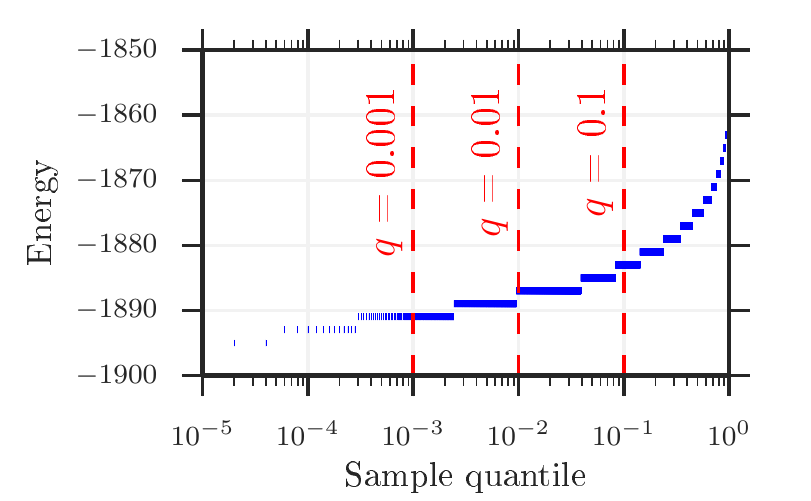}
\caption{\label{fig:ttt-percentiles}Inverse CDF (cumulative distribution function) of the sample energy distribution of the D-Wave hardware.  Vertical lines indicate quantiles used to define target energies.}
\end{minipage}
\hfill
\begin{minipage}[b]{0.48\linewidth}
\centering
\includegraphics[width=\linewidth]{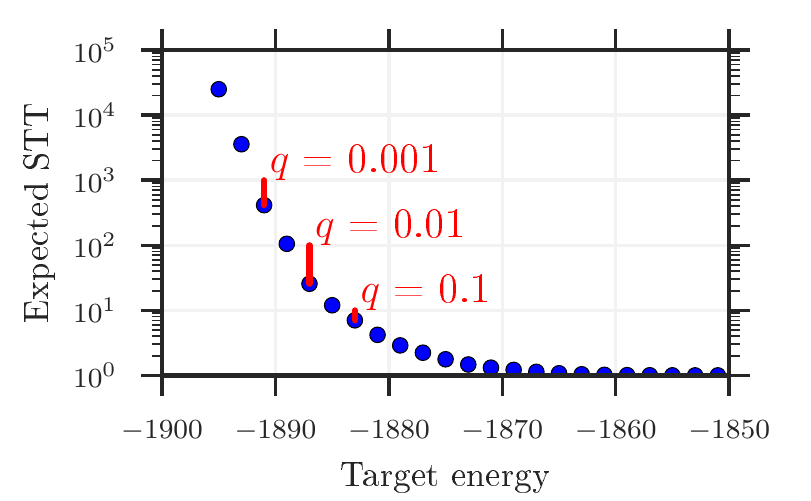}
\caption{\label{fig:ttt-samples}Expected D-Wave STT values for different target energies.  Red line segments indicate the disparity between $1/q$ and expected STT.\\}
\end{minipage}
\end{figure}

\paragraph{Example}
Figure \ref{fig:ttt-percentiles} shows the quantiles (inverse cumulative distribution function) of the sample energy distribution of 50,000 reads from the DW2X on a single input.   We define three target energies for quantiles $q = 0.001, 0.01,  0.1$,  respectively, corresponding to the best  0.1 percent, 1 percent,  and 10 percent of energies in the sample. 

It is natural to intuit that, for a given quantile $q$,  STT would simply be $1/q$.  However, $1/q$ only serves as 
an upper bound on STT:  due to quantization of energy levels the actual STT can be much lower than this bound.  
For example in Figure \ref{fig:ttt-percentiles} the energy -1890  found at $q=0.001$  (suggesting STT = 1000) is also found at $q = 0.002$ and higher quantiles, which means that STT $\leq$ 500 for this input.       

In Figure \ref{fig:ttt-samples} we plot what is essentially the inverse of Figure \ref{fig:ttt-percentiles}.  The expected number of samples required by the D-Wave hardware to hit a given target energy is the reciprocal of the probability mass function at the target energy.  In this example, for quantiles 0.001, 0.01, and 0.1, the STT values are  413.2, 25.58, and 7.029,  lower than their upper bounds of 1000, 100, and 10.  

\paragraph{}
TTT is the expected time required by the solver to reach the target energy, found by combining STT with computation times.  We consider two versions of this metric: total time (\ttttotal)  and annealing only (\tttanneal).   
The two versions serve different purposes in our analyses.  \ttttotal{} measures the entire computation start to finish, including chip programming and sample times.  This  a more realistic  metric that  corresponds to the wall clock time a user would experience.  \tttanneal{}, on the other hand,  includes only the anneal time needed to solve the problem---this more precisely captures the fundamental performance properties of the algorithms considered here, and allows a better view of scaling performance.   Note that I/O times dominate computation costs in current processor models;  this overhead cost has 
been significantly reduced compared to past generations of the processor and is expected to continue to decrease with future releases.

Timing is broken down into \emph{programming/initialization time}, \emph{anneal time}, and \emph{readout time}, respectively $t_i$, $t_a$, and $t_r$.  For the DW2X these are defined in the previous section.  The software solvers used in our comparison study are also designed to return a number of samples for each input; for these, $t_i$ corresponds to initialization/constructor time, $t_a$ corresponds to the bare-bones time to generate one sample, and $t_r$ is considered negligible and recorded as 0 (see Appendix \ref{app:time} for details).  Let $p_t$ be the probability that a single sample will reach the target energy, as estimated by the sample statistic $\hat{p}_t$, which is equal to $1/\text{STT}$. The use of gauge transformations for the DW2X requires us to also consider $p_g$, the probability that a single gauge transformation will contain a successful sample, as estimated by the sample statistic $\hat{p}_g$ (see \cite{Boixo2013} for a description of gauge transformations).  Each gauge transformation requires an additional programming step, adding time $t_i$ to total computation time.  For software solvers we define $p_g=1$.  With these in hand, we have:
\begin{align*}
\text{STT} &= 1/\hat{p}_t \\
\text{\tttanneal} &= t_a/\hat{p}_t \\
\text{\ttttotal} &= \frac{t_i}{\hat{p}_g} + \frac{t_a + t_r}{\hat{p}_t}~.
\end{align*}  See Appendix \ref{app:time} for details of our time measurement procedures.  

Thus,  for  given $q$ the TTT metric identifies as a target the lowest solution energy the D-Wave 2X  processor 
can expect to find within its first  $\text{STT}  \leq  1/q$ samples.  In this study we use quite small 
quantiles $q = 0.001, 0.01, 0.1$  which correspond to sample size bounds of  $1000, 100, 10$:    
the D-Wave 2x finds these samples using total computation times of 352ms,  46ms,  and 15ms at most.  

The TTT comparison essentially asks how much time competing software solvers need to find solution energies of quality 
 that matches the target energies found by D-Wave 2X processor within very short time limits (from  15ms to 352ms).

\section{Experimental setup}
\label{setup} 
 
The goal of this study is to illustrate some properties of the TTT metric by evaluating the performance of the latest generation D-Wave processor on Chimera-structured Ising problems.  This section outlines our experimental procedure.

\subsection{Problems}\label{sec:problems}
The problem classes we consider here fall into three categories:

\paragraph{RAN$\bm{r}$}
In RAN$r$ problems, for integer $r$, each $h$ value is 0 and each $J$ value is chosen uniformly from the integers in the range $[-r, r]$, excluding 0.  We consider classes RAN1, RAN3, RAN7, and RAN127.

\paragraph{AC$\bm{k}$-odd}
The AC$k$ problem class can be generated by taking the RAN1 problem class and multiplying each inter-tile coupling by $k$.  These problems appear to be more challenging than simple random problems by removing a type of  ``local optimality''  that is otherwise created within each tile.  We consider AC3 problems, in which each in-tile coupling is chosen uniformly from $\pm 1$ and each inter-tile coupling is chosen uniformly from $\pm 3$.  The AC3-odd problem class additionally adds fields to reduce degeneracy---for each spin whose incident couplings have an even sum, we add a field chosen uniformly from $\pm 1$.

\paragraph{FL$\bm{r}$}
The class of frustrated loop problems with bounded precision $r$ defined by King \cite{King2015} as a modification of the original idea of Hen et al.~\cite{Hen2015}.  These are constraint satisfaction problems generated for a specific input graph wherein the Hamiltonian range is limited to $r$.  We consider FL3 problems generated with a constraint-to-qubit ratio of 0.25, the hardest regime for these problems.

\paragraph{}
For each problem class tested we generated 100 inputs per problem size over a range of problem sizes from $C_4$ (128 qubits) up to $C_{12}$ (1097 qubits).  

\subsection{Solvers}  
The reference solver, and our primary interest in this study,  is a D-Wave 2X  quantum annealing processor currently online at D-Wave headquarters.  For comparison we use representatives from the two fastest known families of classical software solvers for Chimera-structured Ising problems:  simulated annealing (SA) and the Hamze-de Freitas-Selby algorithm (HFS).  

Each solution from each solver is post-processed by applying a simple greedy descent procedure (greedily flipping spin values) to move the solution to a local minimum.  This removes some noise from the metrics and for these problem classes can be done very quickly.  It also makes our parameterization of simulated annealing more robust, reducing the impact of final temperatures that are too warm.

Throughout, for each input, we use software times corresponding to the {\em fastest}  (best-tuned) version available from each software family.  This essentially assumes the existence of a hypothetical clairvoyant portfolio solver that instantly chooses the best parameter settings for the input and knows exactly how long to run the solver.  We have not similarly tuned the D-Wave processor for best performance;  instead we use default parameter settings 
for all  problems.  

We note that this portfolio approach is somewhat contrary to accepted benchmarking guidelines (\cite{barr1995designing,johnson2002theoretician}), which recommend against too-aggressive tuning and {\em post hoc} selection of solution methods.  Because of this choice the software times reported here correspond to an optimistic usage scenario that is unlikely to arise in practice.  However,  overestimating DW2X times using untuned performance data, and underestimating software times using unrealistically tuned data, allows us to find a lower bound on the size of the performance gap that would  be observed by a user.  See Appendix \ref{app:time} for more discussion of time measurement and Appendix \ref{app:tuning} for our procedures for tuning SA.

\paragraph{D-Wave 2X system}
The DW2X is the latest generation of D-Wave quantum annealer, launched in summer 2015.   We use a fixed annealing time of 20\us{} and fixed rethermalization time of 100\us{} throughout.   Post-processing (greedy descent) is applied in all cases.  Our calculation of \ttttotal{} assumes that gauge transformations might be applied and includes the extra programming cost when appropriate.

\paragraph{Simulated annealing}
Simulated annealing \cite{Kirkpatrick1983} is a canonical optimization algorithm that simulates thermal annealing, the classical analog to quantum annealing.  We use an implementation of simulated annealing developed in-house.  This version of simulated annealing is optimized for Chimera architectures and accepts Hamiltonians stored as single-precision floating point numbers.  

Following R{{\o}}nnow et al.~\cite{Roennow2014} we carried out extensive pilot tests to find optimal parameter settings for (in-house) SA, and report the lowest computation time observed for each input and metric.  See Appendix \ref{app:tuning} for details.  While our implementation of SA should be as good as any other in terms of solution quality, it is not the fastest.  Isakov et al.~\cite{Isakov2014} developed highly optimized simulated annealing codes---one particularly effective technique they use is the recycling of randomly generated numbers; this leads to a significant speedup at the cost of introducing a small amount of correlation between the samples.  Two of their specific SA codes are of interest in this study: \texttt{an\_ms\_r1\_nf} (for RAN1 problems) and \texttt{an\_ss\_ge\_fi} (for other problem classes).  The \texttt{an\_ms\_r1\_nf} solver only works for one class of problems---RAN1 problems---but achieves a huge speed increase by exploiting word-level parallelism to run 64 replicas at a time using bitwise arithmetic.  

We estimate TTT metrics for these two solvers by using the STT results of our in-house SA, then converting these STT results to TTT results using the single-threaded timings reported by Isakov et al.~\cite{Isakov2014}, which were also measured on an Intel Xeon E5-2670.  For all TTT timings reported for SA in this study, we report the TTT value for whichever time is best for the specific input: the measured time of our in-house SA, the estimated time of \texttt{an\_ss\_ge\_fi}, and, if applicable, the estimated time of \texttt{an\_ms\_r1\_nf}.  Again, this is with STT determined using the optimal number of sweeps and better cooling schedule, selected \textit{a posteriori} on a per-input basis.

We note that, according to accepted algorithm benchmarking guidelines \cite{barr1995designing,johnson2002theoretician}, a ``fair test''  requires that solvers used in a given study should be matched by scope; in particular every solver should be able to read and solve every input.  Therefore \texttt{an\_ms\_r1\_nf} should arguably be disqualified from this study because it can only read one of the 6 input classes used.  Nevertheless we include it in the RAN1 tests to identify the extreme boundaries of single-threaded software capabilities.

\paragraph{HFS algorithm}
We use a slightly modified version of Selby's implementation \cite{Selby2013git} of the HFS algorithm \cite{Selby2014, Hamze2004}.  This algorithm performs repeated optimizations on low-treewidth induced subgraphs of the input graph.  It performs a local neighborhood search like SA, but with a much larger neighborhood, and thus slower but more effective updates.  Selby's solver can run in different modes depending on the treewidth of the subgraphs to optimize over.  For this study we tried the treewidth 4 solver (GS-TW1) and the treewidth 8 solver (GS-TW2).  While GS-TW2 shows better performance in finding ground states for large problems \cite{Selby2014}, GS-TW1 showed better time-to-target performance in every case in this study; for this reason we only report results and timings from GS-TW1.

In its original form, Selby's implementation searches repeatedly for minima and discards any local optima that turn out to not be globally  optimal;   his timing procedure also ignores some initial ``burn in''  time that is spent finding non-optimal solutions.  We use the solver in what we call ``sampling mode'', which returns all local minima  found as independent samples and records time-per-sample for each,
which corresponds to anneal times for SA and DW2X.  Since we do not discard local minima, this version of the solver is faster with respect to the TTT metric.  

\subsection{Metrics}  

To define target energies for each input  we drew 50,000 samples from the D-Wave 2X system, taking
 1,000 samples from each of 50 random gauge transformations.  Each gauge transformation slightly perturbs the nominal Hamiltonian:  this reduces bias and creates a more accurate reference distribution from which sample quantiles may be drawn.  
We find  target energies for  quantiles $q=$ 0.001, 0.01, and 0.1,  which give  {\em  upper bounds}  on  STT of 1000, 100 and  10 hardware samples, respectively.   For the DW2X,  these sample counts correspond respectively to total computation times of  352ms, 46ms, and 15ms,  and to anneal times of  20ms,  2ms, and 0.2ms.   

To calculate  STT for each input, we used the same 50,000 samples from the DW2X.  For  HFS and each parameterized version 
of SA we calculated STT based on 1,000 samples for each input;   the use of smaller sample sizes creates coarser increments of sample counts STT in these cases,  but this step was necessary given the large computation times required by software 
to return 1000 samples; see Appendix A for details.     

\section{Results}
\label{results}  

Our results for the full suite of input classes appear in Appendix \ref{app:results}.   Here we show selected examples that highlight the range of relative performance advantages of the D-Wave 2X processor.

\begin{figure}
\centering
\includegraphics[width=\textwidth]{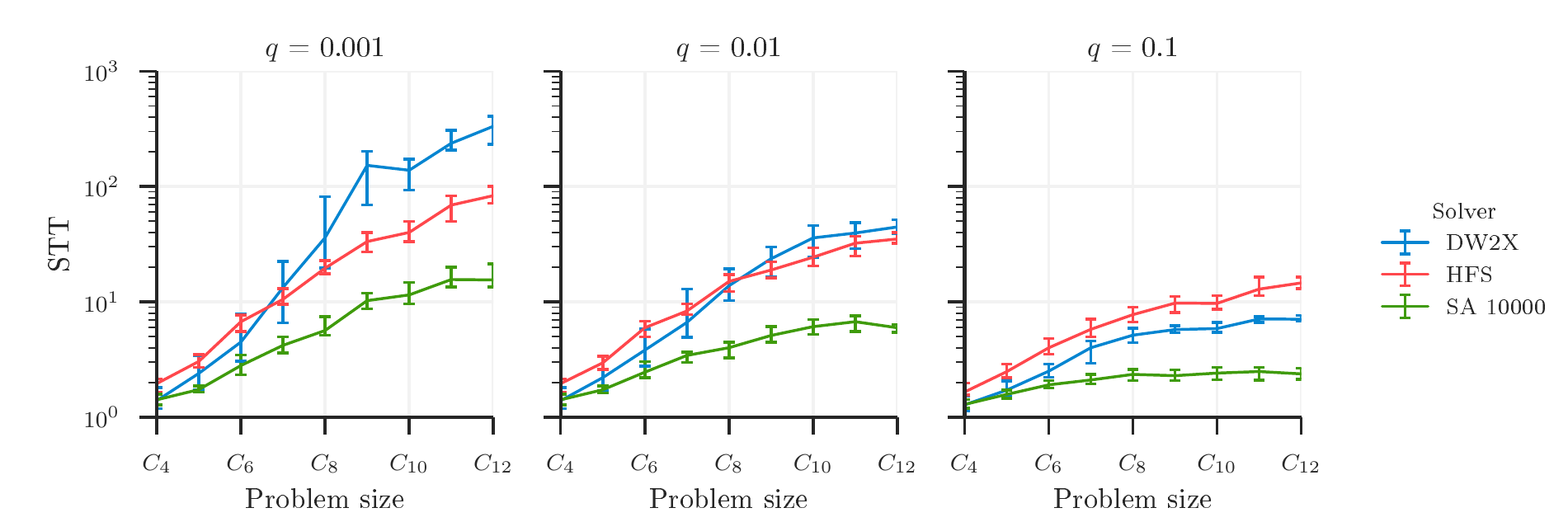}
\caption{\label{fig:ran1-sample} Performance of solvers in STT for RAN1 problems.  Shown are the medians (over 
100 input instances)  for each solver and problem size, with error bars showing 95\% confidence intervals.  SA performance corresponds to SA10000 (10000 sweeps per anneal) which is optimal in this metric.} 
\end{figure}  

\begin{figure}
\centering
\includegraphics[width=\textwidth]{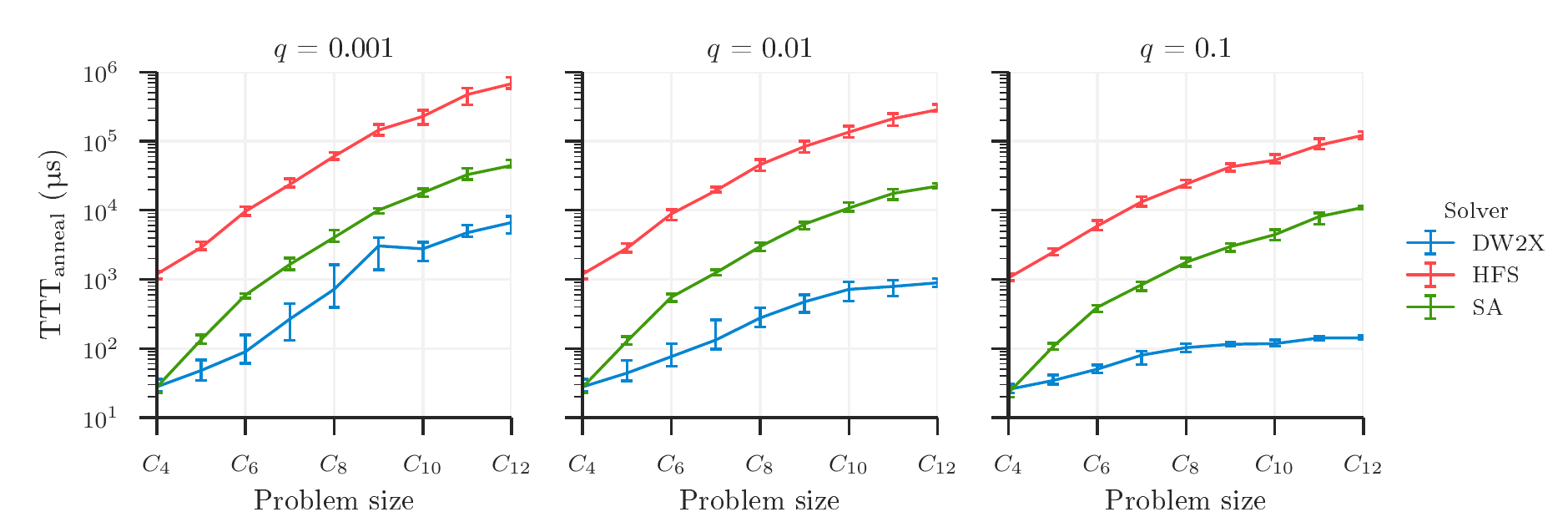}
\caption{\label{fig:ran1-anneal}Performance of solvers in \tttanneal{} for RAN1 problems.  Shown are medians for each solver and size with error bars showing 95\% confidence intervals.  SA times are estimated for the   \texttt{an\_ms\_r1\_nf} solver; see Section \ref{discussion} for details.  }
\end{figure}

\begin{figure}
\centering
\includegraphics[width=\textwidth]{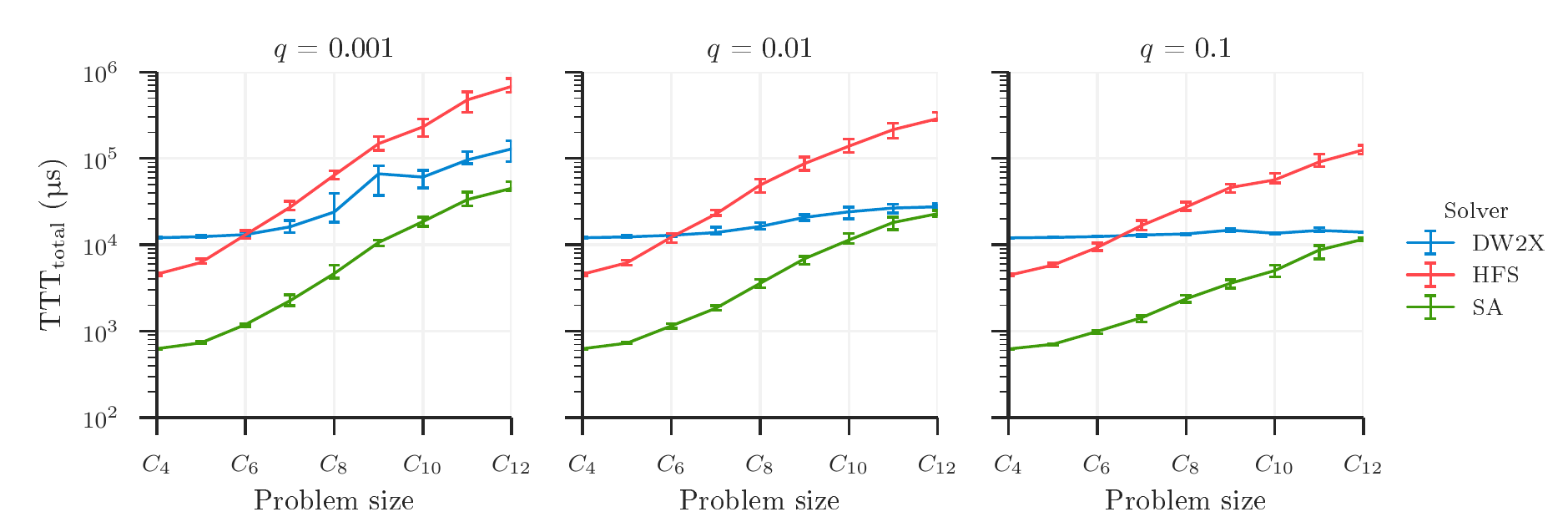}
\caption{\label{fig:ran1-total}Performance of solvers in \ttttotal{} for RAN1 problems.  Shown are medians for each solver and size with error bars showing 95\% confidence intervals.  For this input class, SA times are estimated and correspond  to the  \texttt{an\_ms\_r1\_nf} solver described in \cite{Isakov2014}; see Section \ref{discussion} for details. }  
\end{figure}

Figure \ref{fig:ran1-sample} shows median STT performance (over 100 inputs) and 95\% confidence intervals\footnote{Throughout this paper we report confidence intervals for medians derived analytically from the sample quantiles (see, e.g., \cite{hollander2013nonparametric}).  For 100 inputs the 95\% CI for the median is bounded by the 40th and 60th percentiles.} for RAN1 problems.  Left to right,  the three panels correspond to $q=0.001$ (harder targets),  $q=0.01$, and $q=0.1$ (easier targets).   The $x$-axis shows increasing Chimera sizes $C_4$ to $C_{12}$; recall that the number of qubits (variables) in a $C_s$ problem is approximately $8s^2$. The $y$-axis (log scale) shows expected STT.  For  DW2X,  the blue curves correspond to sample counts bounded above by 1000, 100, and 10, for increasing quantiles $q$:  here we see STS values of approximately  300,  40, and  8, respectively at $C_{12}$ size.  The confidence intervals are small, corresponding to a narrow range of observations over all inputs.  By definition, the set of possible STT values for the DW2X are bounded above by 1000, 100, and 10, which are quite close to the  median values shown at largest problem sizes.  We observe similarly tight ranges for the software solvers, and there is no reason to expect heavy tails as observed by Steiger et al.~\cite{steiger2015heavy} in the STS metric.

The fact that STT for the reference solver (DW2X) tends to grow with problem size rather than staying constant appears to be an artifact of the discretized nature of the solution space and the fact that smaller problems are easier to solve.  For example, for $q=0.001$ at  $C_{12}$  size,  the proportion of samples that hit the target is $0.003$; at $C_8$ the proportion is a much higher $0.025$---the smaller problems have fewer distinct energies to choose from as well as a higher relative number of good-quality solutions in the sample.  STT for the reference solver approaches the upper bound of 1/$q$ much more quickly for high-precision inputs than for low-precision inputs.

Among SA solvers we tested,  SA 10000 (using the maximum number of sweeps per sample that we tested) is always optimal for STT because this metric essentially assumes constant cost per sample:  since  more sweeps are free it is always best to maximize the number of sweeps per sample.  The  comparatively flat scaling for SA in this figure indicates that 10000 sweeps is more than ample time to solve these problems:  at $q=0.1$ we see  STT $\leq 2$ throughout, indicating that 50 percent of the samples hit this target.  

It is interesting that the  DW2X and SA curves have approximately the same shape in these three charts, suggesting that the distributions of solutions returned by these solvers are quite similar,  at least for quantiles $q=0.1$ and smaller.   The HFS algorithm shows relatively different behavior indicating a different distribution:  worse than DW2X and SA at hitting targets in the smallest quantile (harder problems) but better at hitting targets in the biggest quantile (easier problems).  

Although SA 10000 is best in STT,  the  large time-per-sample it requires (approximately 0.16\us{} per sweep or 1.6ms per sample at $C_{12}$ size), means that it is not necessarily  the best choice for the TTT metric.   Figure \ref{fig:ran1-anneal} shows \tttanneal{} for these problems, calculated by multiplying STT by the time-per-anneal  for each solver.  We have the following observations:  

 \begin{itemize}  
\item For DW2X,  time per anneal is a constant 20\us{}.  Therefore scaling of \tttanneal{} is identical to that for STT:   for example both STT and \tttanneal{} increase through about 2.5 orders of magnitude at $q=0.001$,  and increase about 3-fold at $q=0.1$.   

\item For SA,  time per anneal is the product of  time per sweep and the number of sweeps per anneal.   Time per sweep is proportional to the number of problem variables and grows quadratically in $C_s$.  In  this and subsequent figures  we display results for the {\em optimal} number of sweeps found in extensive pilot tests of a variety of solver parameters.  For RAN1 inputs  the best version of SA corresponds to \texttt{an\_ms\_r1\_nf} and the runtimes shown  are based on estimates from data published elsewhere. See Appendix \ref{app:time} for details about software times and Appendix \ref{app:tuning} for our tuning procedures for SA.

Comparing the green curves in Figures \ref{fig:ran1-sample} and \ref{fig:ran1-anneal} we see that STT grows by factors near 10x, 10x, and 2x in each panel,  while TTT grows by about 1000x,  1000x,  and 100x,  respectively,  as problem size increases about 8-fold from $C_4$ (about 128 variables)
to  $C_{12}$ (1097) variables.  Thus, roughly speaking, a 1000x fold increase in computation time is a combination of an 8-fold increase in time-per-sweep  and a 125-fold increase in sweeps-per-solution  (proportional to STT for this particular version of SA) through the problem range.  

\item For HFS, time per anneal varies depending on how long it takes to hit a local minimum.  In this figure TTT scaling resembles that of SA.  

\item  In all cases,  at $C_{12}$ sizes, \tttanneal{} is between 8x and about 80x faster than the best competing software solvers we know of.  

\item Although the scaling curves shown here give  a good idea of the relative amount of work  each solver performs in terms of computation time,  we caution against trying to extend these curves to larger problem sizes or other quantiles.  For example the shapes of the curves seen here are  an  artifact of how this choice of quantiles ``cuts''  the distribution solution energies, and would be different, for example, if the target were selected by fixing computation times.  Much more work is needed to understand the tradeoff between computation time and solution quality that is offered by each solver; this is an interesting  topic for future research.   
\end{itemize}

Figure \ref{fig:ran1-total} shows \ttttotal{} for RAN1 problems;  this metric includes initialization time and (in the case of DW2X)  readout time and reflects the actual user experience using these solvers.  The flat scaling of the  blue DW2X curve on the center and right panels, and at small $C_s$ values on the left panel, reflects the fact that the constant  programming time of 11.6ms dominates total computation time.   Programming time dominates whenever the number of samples is less than 36;  in particular for  $q=0.1$ the upper bound of 10 reads (3.4ms) ensures that this curve will always appear flat (see the discussion of component times in Section \ref{sec:overview}).  
Only at largest problems and at the hardest quantile ($q=0.001$) do we see computation times grow above this threshold.  

In all three quantiles DW2X is faster in total computation time than the HFS algorithm at the largest problem sizes.  In all three quantiles the SA solver is fastest; estimated \texttt{an\_ms\_r1\_nf} times are lowest among all solvers considered, at all problem sizes.  Recall that this solver exploits bit parallelism to achieve extremely fast computation times at the price of extremely narrow applicability to the 1-bit problems tested here. 

\begin{figure}
\centering
\includegraphics[width=\textwidth]{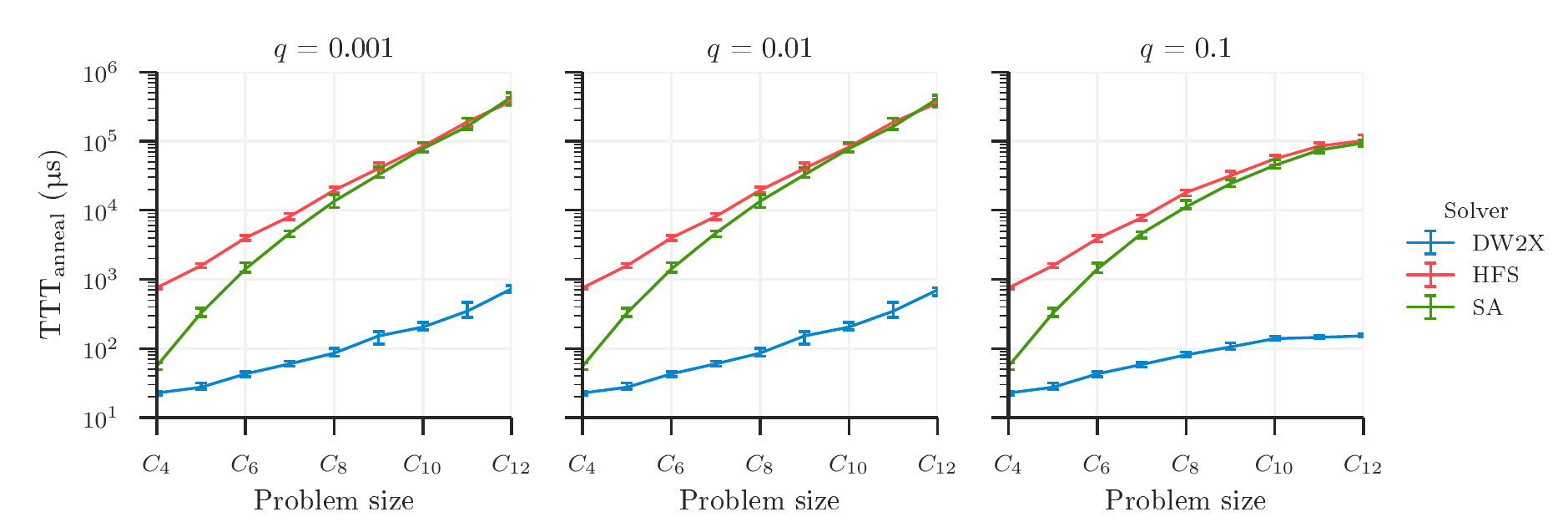}\caption{\label{fig:fl3-anneal}Performance of solvers in \tttanneal{} for frustrated loop (FL3) problems.  Shown are medians over 100 random inputs for each solver and size with error bars showing 95\% confidence intervals.}
\end{figure}

\begin{figure}
\centering
\includegraphics[width=\textwidth]{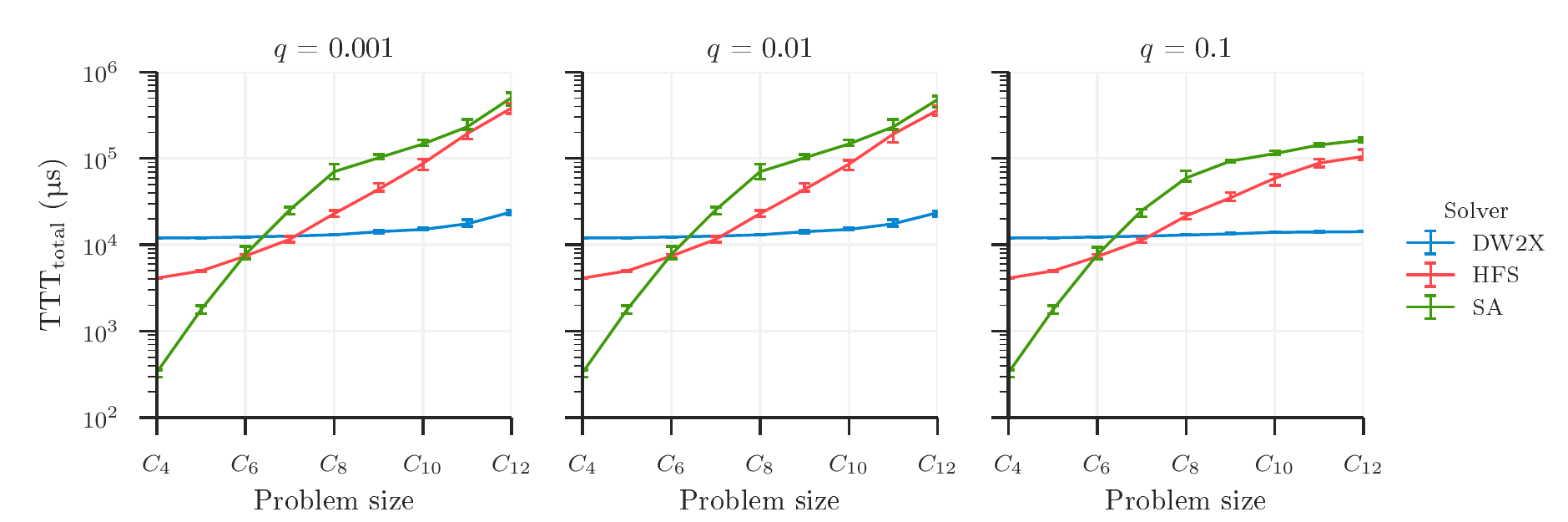}\caption{\label{fig:fl3-total}Performance of solvers in \ttttotal{} for frustrated loop (FL3) problems.  Shown are medians over 100 random inputs for each solver and size with error bars showing 95\% confidence intervals.}
\end{figure}

Figures 4 through 6 illustrate the strongest relative performance in TTT that we have observed for software solvers.   The next three figures illustrate the other extreme.  In Figures \ref{fig:fl3-anneal} and \ref{fig:fl3-total} we show \tttanneal{} and \ttttotal{}, respectively, for FL3 problems.  For \tttanneal{}, at the largest problem sizes the DW2X is about 600x faster than any competitor for all quantiles measured.  As with RAN1 problems, for the DW2X, the programming time dominates \ttttotal{} in nearly all cases.  STT plots for FL3 and other problem classes can be found in Appendix \ref{app:results}.

\begin{figure}
\centering
\includegraphics[width=\linewidth]{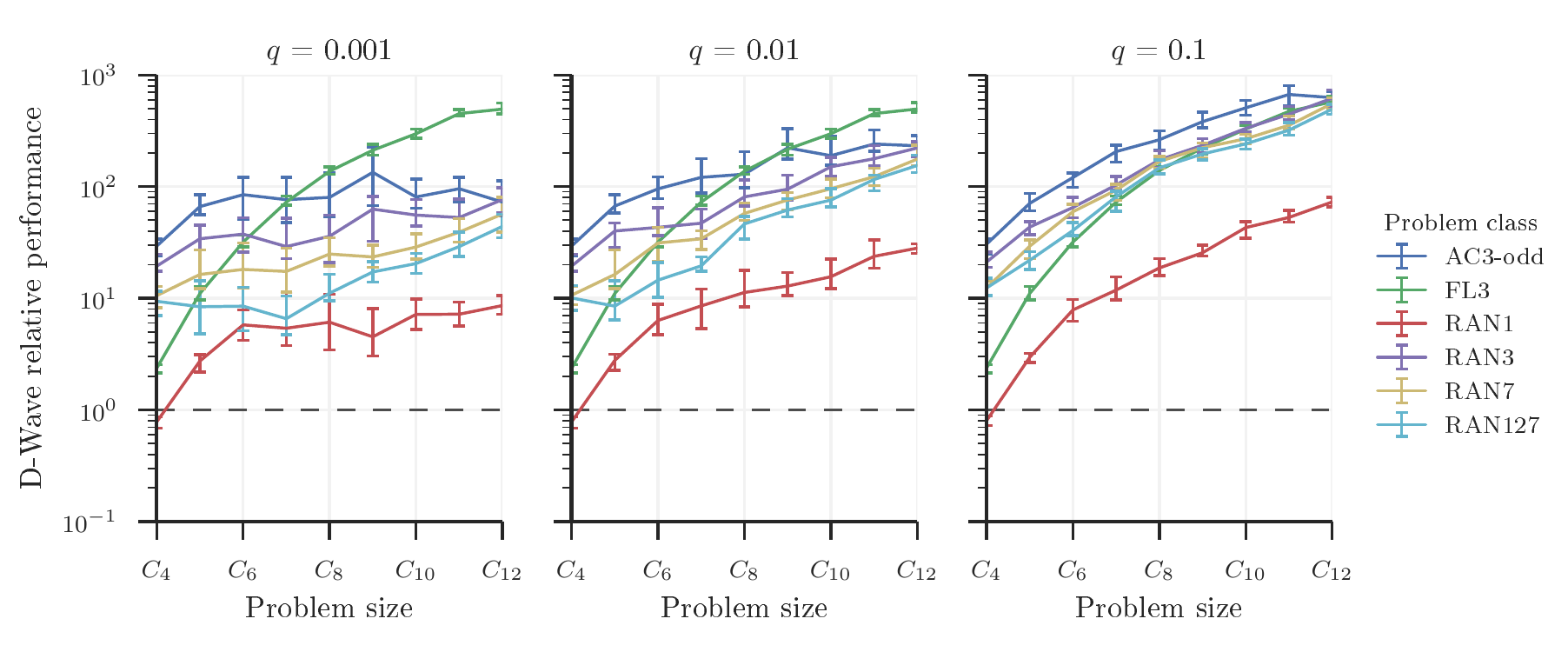}
\caption{\label{fig:advantage_anneal}Relative performance of the DW2X vs.~other solvers in \tttanneal.  Shown are medians for each problem class and size with error bars showing 95\% confidence intervals.}
\end{figure}

\begin{figure}
\centering
\includegraphics[width=\linewidth]{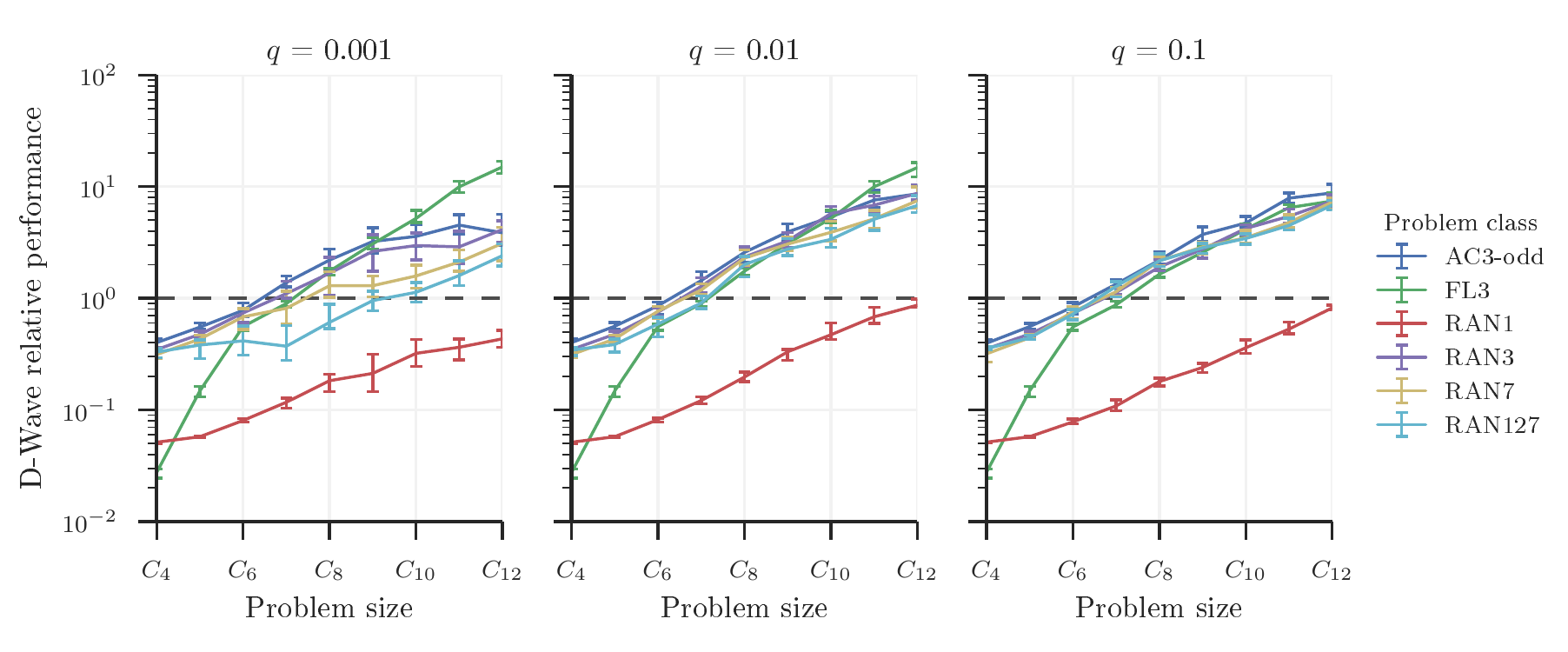}
\caption{\label{fig:advantage_total}Relative performance of the DW2X vs.~other solvers in \ttttotal.  Shown are medians for each problem class and size with error bars showing 95\% confidence intervals.}
\end{figure}

\paragraph{Measuring D-Wave relative performance} By dividing the best software solver's TTT by the TTT of the DW2X for each input, we can measure how much faster or slower the D-Wave processor is than the strongest competitor on a per-input basis.  We plot this relative performance for \tttanneal{} and \ttttotal{} in Figures \ref{fig:advantage_anneal} and \ref{fig:advantage_total} respectively.  

As a function of $q$, the relative performance of the D-Wave 2X system over the software solvers generally peaks at $q=0.1$ for \tttanneal{} and at $q=0.01$ for \ttttotal{}.  This range seems to be the sweet spot for the DW2X, where it has gathered enough samples to ensure a good solution, yet the software solvers have not had enough time to catch up.  For \ttttotal{}, the D-Wave relative performance is notably lower for $q=0.001$; this is partially due to the possibility of requiring multiple gauge transformations to reach the target energy.  It is possible that D-Wave's relative performance in this metric could be improved by increasing the number of samples per gauge transformation.

On RAN$r$ problems there is a steady degradation of SA as a competitor as precision increases (see Appendix \ref{app:results}), even without considering the \texttt{an\_ms\_r1\_nf} solver.  This may be because as precision increases, degeneracy decreases, meaning the problems are harder for all solvers.  It may also be because the cooling schedules we use for SA are inappropriate for higher-precision instances. While we have made a good-faith effort to use SA optimally, in particular by using reasonable cooling schedules and optimizing the number of sweeps as suggested by R{{\o}}nnow et al.~\cite{Roennow2014}, it is possible that different parameterization could lead to better performance, and we invite others to try.  In the meantime, for high-precision random instances we are content to let HFS stand as the most competitive software solver.

\section{Discussion}
\label{discussion}

In these TTT metrics, with the exception of the RAN1 problem class, the single-threaded software solvers evaluated here have not kept up with the D-Wave hardware at the full 1097-qubit problem scale.  While it is possible that a new algorithm could be developed that could beat the DW2X in these metrics on a single thread, and we encourage researchers to continue such efforts, we believe that the real question has now turned to multithreaded, multi-core software solvers.  

Isakov et al.~\cite{Isakov2014} evaluated parallelized versions of their simulated annealing code with up to 16 threads, but this would be insufficient to match the anneal-time-only performance of the DW2X in many cases, even with idealized perfect parallelism.  
For these TTT metrics it remains unclear how many CPU cores it would take to match the performance of the DW2X.  It bears investigating this question using actual timing on multi-CPU platforms, where memory access and communication costs are likely to dominate single-core instruction times at least as much as programming and readout times dominate the quantum annealer. 

While this study has only used CPU-based software solvers, GPUs are becoming increasingly popular as sources of cheap parallelism and are a viable means to fast, cheap Monte Carlo simulation.  We are currently investigating GPU-based algorithms to determine how many GPU cores it would take to match the DW2X in these TTT metrics.

\bibliography{paper}

\begin{thebibliography}{10}

\bibitem{Harris2010}
R.~Harris, M.W. Johnson, T.~Lanting, A.J. Berkley, J.~Johansson, P.~Bunyk,
  E.~Tolkacheva, E.~Ladizinsky, N.~Ladizinsky, T.~Oh, et~al.
\newblock Experimental investigation of an eight-qubit unit cell in a
  superconducting optimization processor.
\newblock {\em Physical Review B}, 82(2):024511, 2010.

\bibitem{johnson2011quantum}
M.W. Johnson, M.H.S. Amin, S.~Gildert, T.~Lanting, F.~Hamze, N.~Dickson,
  R.~Harris, A.J. Berkley, J.~Johansson, P.~Bunyk, et~al.
\newblock Quantum annealing with manufactured spins.
\newblock {\em Nature}, 473(7346):194--198, 2011.

\bibitem{Boixo2013}
S.~Boixo, T.~Albash, F.M. Spedalieri, N.~Chancellor, and D.A. Lidar.
\newblock Experimental signature of programmable quantum annealing.
\newblock {\em Nature communications}, 4, 2013.

\bibitem{barr1995designing}
R.S. Barr, B.L. Golden, J.P. Kelly, M.G.C. Resende, and W.R. Stewart~Jr.
\newblock Designing and reporting on computational experiments with heuristic
  methods.
\newblock {\em Journal of Heuristics}, 1(1):9--32, 1995.

\bibitem{bartz2010experimental}
T.~Bartz-Beielstein, M.~Chiarandini, L.~Paquete, and M.~Preuss.
\newblock {\em Experimental methods for the analysis of optimization
  algorithms}.
\newblock Springer, 2010.

\bibitem{jain2008art}
R.~Jain.
\newblock {\em The art of computer systems performance analysis}.
\newblock John Wiley \& Sons, 2008.

\bibitem{johnson2002theoretician}
D.S. Johnson.
\newblock A theoretician’s guide to the experimental analysis of algorithms.
\newblock {\em Data structures, near neighbor searches, and methodology: fifth
  and sixth DIMACS implementation challenges}, 59:215--250, 2002.

\bibitem{parekh2015benchmarking}
O.~Parekh, J.~Wendt, L.~Shulenburger, A.~Landahl, J.~Moussa, and J.~Aidun.
\newblock Benchmarking adiabatic quantum optimization for complex network
  analysis.
\newblock Technical Report SAND2015-3025, Sandia National Laboratories, 2015.

\bibitem{McGeoch2013}
C.C. McGeoch and C.~Wang.
\newblock Experimental evaluation of an adiabiatic quantum system for
  combinatorial optimization.
\newblock In {\em Proceedings of the ACM International Conference on Computing
  Frontiers}, page~23. ACM, 2013.

\bibitem{Boixo2014}
S.~Boixo, V.N. Smelyanskiy, A.~Shabani, S.V. Isakov, M.~Dykman, V.S. Denchev,
  M.~Amin, A.~Smirnov, M.~Mohseni, and H.~Neven.
\newblock Computational role of collective tunneling in a quantum annealer.
\newblock {\em arXiv preprint arXiv:1411.4036}, 2014.

\bibitem{Hen2015}
I.~Hen, J.~Job, T.~Albash, T.F. R{\o}nnow, M.~Troyer, and D.~Lidar.
\newblock {Probing for quantum speedup in spin glass problems with planted
  solutions}.
\newblock {\em arXiv preprint arXiv:1502.01663v2}, 2015.

\bibitem{Venturelli2014}
D.~Venturelli, S.~Mandr{\`a}, S.~Knysh, B.~O'Gorman, R.~Biswas, and
  V.~Smelyanskiy.
\newblock Quantum optimization of fully-connected spin glasses.
\newblock {\em arXiv preprint arXiv:1406.7553}, 2014.

\bibitem{Roennow2014}
T.F. R{{\o}}nnow, Z.~Wang, J.~Job, S.~Boixo, S.V. Isakov, D.~Wecker, J.M.
  Martinis, D.A. Lidar, and M.~Troyer.
\newblock Defining and detecting quantum speedup.
\newblock {\em Science}, 345(6195):420--424, 2014.

\bibitem{Selby2014}
A.~Selby.
\newblock {Efficient subgraph-based sampling of Ising-type models with
  frustration}.
\newblock {\em arXiv preprint arXiv:1409.3934v1}, 2014.

\bibitem{Selby2013git}
A.~Selby.
\newblock {QUBO-Chimera}.
\newblock \url{https://github.com/alex1770/QUBO-Chimera}, 2013.
\newblock GitHub repository.

\bibitem{Hamze2004}
F.~Hamze and N.~de~Freitas.
\newblock From fields to trees.
\newblock In {\em Proceedings of the 20th conference on Uncertainty in
  artificial intelligence}, pages 243--250. AUAI Press, 2004.

\bibitem{cplex2015}
IBM~ILOG CPLEX.
\newblock Optimization studio 12.6.2, 2015.

\bibitem{impagliazzo1999complexity}
R.~Impagliazzo and R.~Paturi.
\newblock Complexity of {$k$}-sat.
\newblock In {\em Computational Complexity, 1999. Proceedings. Fourteenth
  Annual IEEE Conference on}, pages 237--240. IEEE, 1999.

\bibitem{King2015}
A.D. King.
\newblock {Performance of a quantum annealer on range-limited constraint
  satisfaction problems}.
\newblock {\em arXiv preprint arXiv:1502.02098v1}, 2015.

\bibitem{McGeoch2012}
C.C. McGeoch.
\newblock {\em A guide to experimental algorithmics}.
\newblock Cambridge University Press, 2012.

\bibitem{Saket2013}
R.~Saket.
\newblock {A PTAS for the classical Ising spin glass problem on the Chimera
  graph structure}.
\newblock {\em arXiv preprint arXiv:1306.6943}, 2013.

\bibitem{Young2013}
K.C. Young, R.~Blume-Kohout, and D.A. Lidar.
\newblock {Adiabatic quantum optimization with the wrong Hamiltonian}.
\newblock {\em Physical Review A}, 88(6):062314, 2013.

\bibitem{Pudenz2014}
K.L. Pudenz, T.~Albash, and D.A. Lidar.
\newblock Error-corrected quantum annealing with hundreds of qubits.
\newblock {\em Nature Communications}, 5, 2014.

\bibitem{Vinci2014}
W.~Vinci, T.~Albash, A.~Mishra, P.A. Warburton, and D.A. Lidar.
\newblock {Distinguishing classical and quantum models for the D-Wave device}.
\newblock {\em arXiv preprint arXiv:1403.4228}, 2014.

\bibitem{bunyk2014architectural}
P.~Bunyk, E.~Hoskinson, M.~Johnson, E.~Tolkacheva, F.~Altomare, A.~Berkley,
  R.~Harris, J.~Hilton, T.~Lanting, A.~Przybysz, et~al.
\newblock Architectural considerations in the design of a superconducting
  quantum annealing processor.
\newblock {\em IEEE Transactions on Applied Superconductivity}, 2014.

\bibitem{Dickson2013}
N.G. Dickson et~al.
\newblock {Thermally assisted quantum annealing of a 16-qubit problem}.
\newblock {\em Nature Communications}, 4(May):1903, January 2013.

\bibitem{Lanting2014}
T.~Lanting, A.J. Przybysz, A.Yu Smirnov, F.M. Spedalieri, M.H. Amin, A.J.
  Berkley, R.~Harris, F.~Altomare, S.~Boixo, P.~Bunyk, et~al.
\newblock Entanglement in a quantum annealing processor.
\newblock {\em Physical Review X}, 4(2):021041, 2014.

\bibitem{Farhi2001}
E.~Farhi, J.~Goldstone, S.~Gutmann, J.~Lapan, A.~Lundgren, and D.~Preda.
\newblock {A quantum adiabatic evolution algorithm applied to random instances
  of an NP-complete problem}.
\newblock {\em Science}, 292(5516):472--475, 2001.

\bibitem{Kirkpatrick1983}
S.~Kirkpatrick, C.D. Gelatt~Jr, and M.P. Vecchi.
\newblock Optimization by simmulated annealing.
\newblock {\em Science}, 220(4598):671--680, 1983.

\bibitem{Isakov2014}
S.V. Isakov, I.N. Zintchenko, T.F. R{\o}nnow, and M.~Troyer.
\newblock {Optimized simulated annealing for Ising spin glasses}.
\newblock {\em arXiv preprint arXiv:1401.1084}, 2014.

\bibitem{hollander2013nonparametric}
M.~Hollander, D.A. Wolfe, and E.~Chicken.
\newblock {\em Nonparametric statistical methods}.
\newblock John Wiley \& Sons, 2013.

\bibitem{steiger2015heavy}
D.S. Steiger, T.F. R{\o}nnow, and M.~Troyer.
\newblock Heavy tails in the distribution of time-to-solution for classical and
  quantum annealing.
\newblock {\em arXiv preprint arXiv:1504.07991}, 2015.

\end{thebibliography}

\appendix
\newpage

\section{DW2X hardware working graph}\label{app:hardware}  

\begin{figure}[h!]
\centering
\includegraphics[width=\linewidth]{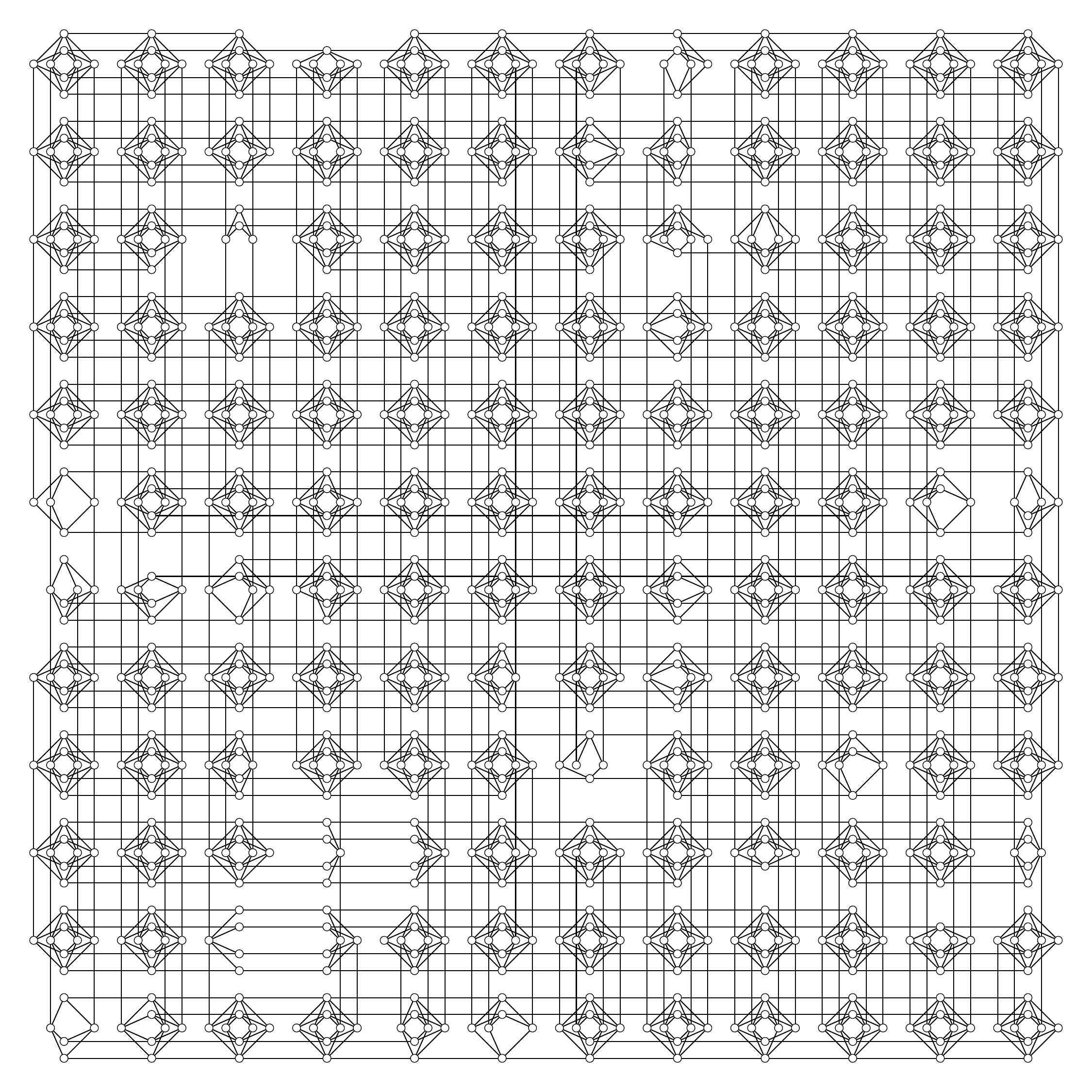}
\caption{\label{fig:bay16}Hardware graph used in the D-Wave 2X system.  The $C_{12}$ Chimera architecture consists of a $12\times 12$ grid of Chimera unit tiles, each having 8 qubits.  Imperfections in the fabrication process lead to inactive qubits; in this chip 1097 of the 1152 qubits are active.}
\end{figure}

\section{Solver timing}
\label{app:time} 

\subsection*{DW2X timing}

The timing for the DW2X does not depend on the size or class of the problem.

\begin{table}[h]
  \begin{center}\begin{small}
    \begin{tabular}{| c | c | c |}
      \hline
      {\bf Programming time} & {\bf Anneal time} &  {\bf Readout time} \\ \hline
	  11.6ms & 20\us & 320\us \\
      \hline
    \end{tabular}\end{small}\vspace{-1em}
\end{center}\caption{Timing components for the D-Wave 2X system.}\label{table:dw_timing}
\end{table}

\subsection*{Software timing}

\begin{figure}
\begin{minipage}[b]{0.48\linewidth}
\centering
\includegraphics[width=\linewidth]{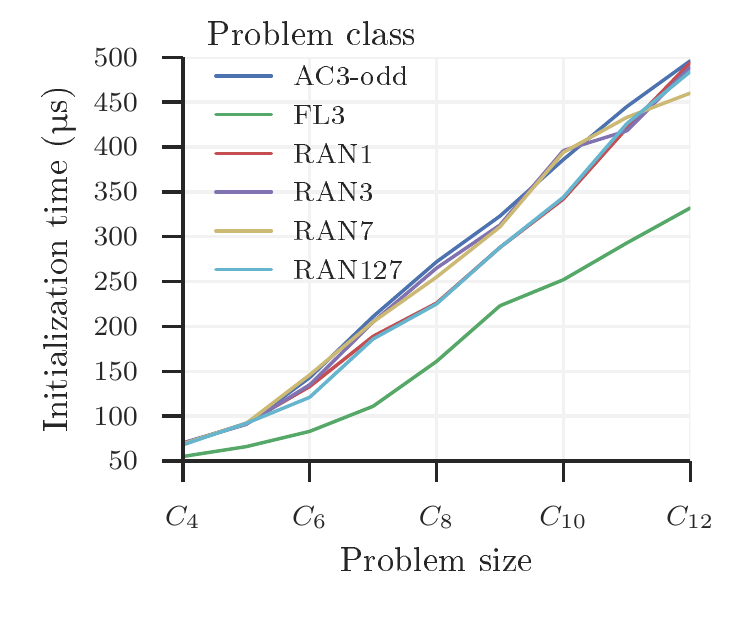}
\caption{\label{fig:sa_time_per_sweep}Initialization time for SA.}
\end{minipage}
\hfill
\begin{minipage}[b]{0.48\linewidth}
\centering
\includegraphics[width=\linewidth]{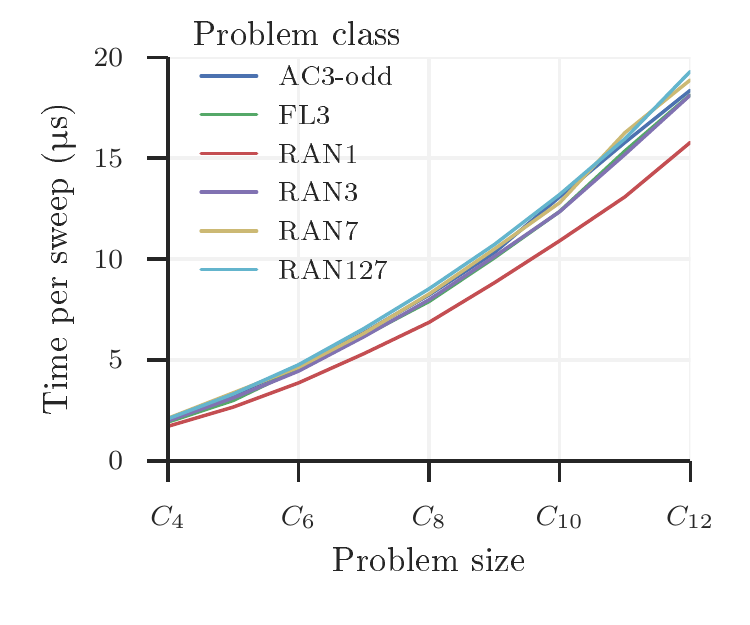}
\caption{\label{fig:sa_init_time}Time per sweep for SA.} 
\end{minipage}
\end{figure}

\begin{figure}
\begin{minipage}[b]{0.48\linewidth}
\centering
\includegraphics[width=\linewidth]{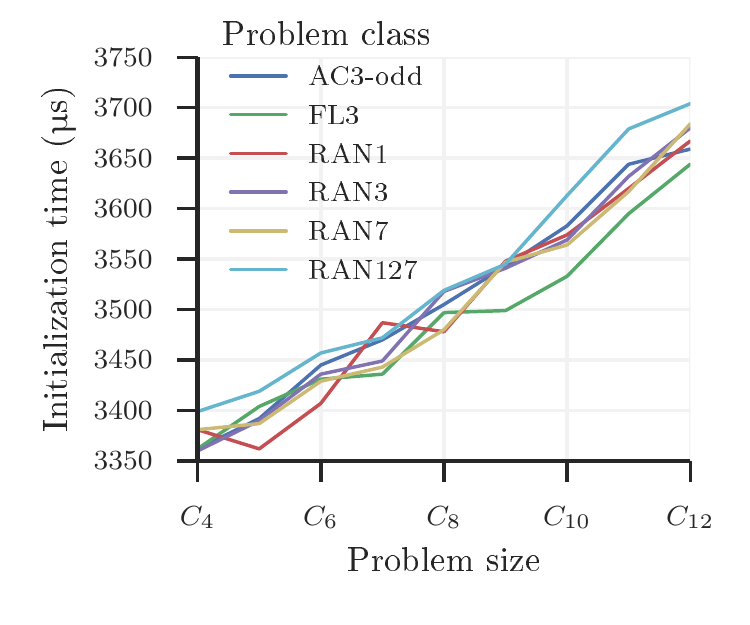}
\caption{\label{fig:hfs_time_per_sweep}Initialization time for HFS.}
\end{minipage}
\hfill
\begin{minipage}[b]{0.48\linewidth}
\centering
\includegraphics[width=\linewidth]{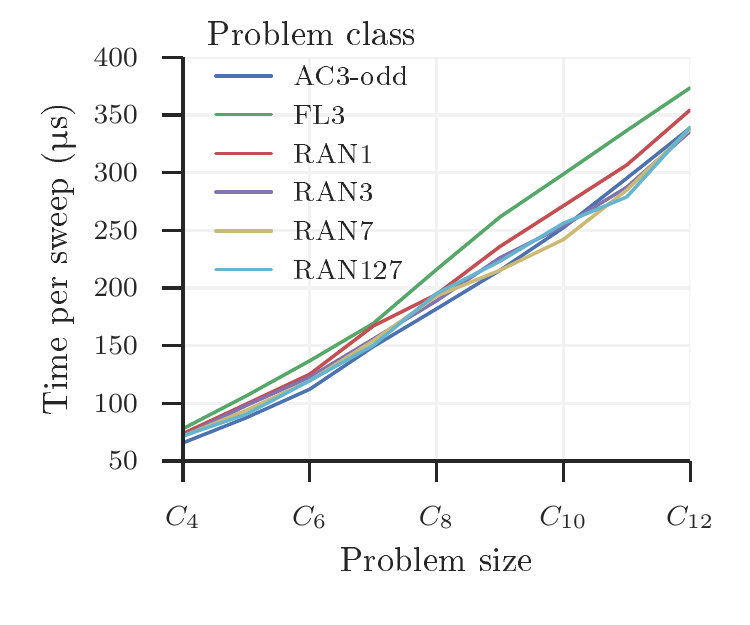}
\caption{\label{fig:hfs_init_time}Time per sweep for HFS.}
\end{minipage}
\end{figure}

Software solvers were run single-threaded on an Intel Xeon E5-2670 processor.  For HFS and our in-house SA implementation we measure initialization time and time per sweep as a function of problem class and size.  To measure initialization time, we timed the initialization method 10 times and took the minimum initialization time.  To measure time per sweep, we measured the time required to perform hundreds of thousands of sweeps and took the mean.  Initialization times and times per sweep for SA and HFS are shown in Figures \ref{fig:sa_time_per_sweep}, \ref{fig:sa_init_time}, \ref{fig:hfs_time_per_sweep}, and \ref{fig:hfs_init_time}.  Note that time per sweep for both solvers grows linearly with the number of spins (and thus quadratically with the Chimera size), as we would expect.  

For the \texttt{an\_ss\_ge\_fi} and \texttt{an\_ms\_r1\_nf} solvers we used initialization times and spin update times reported by Isakov et al.~\cite{Isakov2014} (which were also measured on an Intel Xeon E5-2670).  For the \texttt{an\_ms\_r1\_nf} solver, samples are requested in batches of 64.  For bookkeeping, we define the empirical success probability of a batch hitting the target energy as $\hat{p}_t' = 1 - (1-\hat{p}_t)^{64}$ and the annealing time of a batch as the time required to anneal 64 samples.  A ``spin flip'' corresponds to the cost of updating a single problem variable (a spin) in a single iteration of an inner loop; a ``sweep'' corresponds to a single iteration through all $n$ problem variables.  Table \ref{table:sa_timing} shows initialization time and time per sweep at the largest $C_{12}$ problem size.

\begin{table}[h]
  \begin{center}\begin{small}
    \begin{tabular}{|l|c|c|c|c|}
      \hline
      {\bf Solver} & {\bf Init} & {\bf Spin flips per ns}  & {\bf Time per spin flip } & {\bf Time per sweep (1097 spins)} \\ \hline 
      \texttt{an\_ms\_r1\_nf}  & ~0.6ms  & 6.65    &  0.15ns &  0.16\us{}  \\ 
      \texttt{an\_ss\_ge\_fi}   & 69.0ms & 0.30    &  3.33ns  & 3.66\us{} \\   \hline
    \end{tabular}\end{small}\vspace{-1em}
\end{center}\caption{Timing components for simulated annealing codes of Isakov et al.~\cite{Isakov2014}.}\label{table:sa_timing}
\end{table}

For SA the conversion from sweep time to total time is trivial since the number of sweeps is fixed.  For HFS, on the other hand, the number of ``tree sweeps'' (i.e., low-treewidth updates) per sample is determined internally, as the algorithm simply descends to a local minimum using as many sweeps as it takes.  In this case we measure the mean number of ``tree sweeps''  per sample returned.

\section{Tuning SA}\label{app:tuning} Our implementation of SA has two main parameters: number of sweeps per anneal,  and the choice of cooling schedule.  To optimize the number of sweeps per anneal,  we tried values in the set \{10, 20, 40, 100, 200, 400, 1000, 2000, 4000, 10000\} for all inputs.   We also tried each solver using two cooling schedules:  the first uses an `unscaled' schedule so that the inverse temperature $\beta$ increases linearly from $0.01$ to $3$;  the second uses a `scaled' schedule that increases from $0.01/r$ to $3/r$, where $r$ is the range of the Hamiltonian, i.e., the largest absolute value of any component of the Hamiltonian.  The unscaled version is too fast for high precision inputs (RAN127 and RAN7) and the scaled version was used instead.  For each problem class we report only results from the better of the two cooling schedules and best choice of sweeps per anneal.

%
%
%
%

\section{Full results}\label{app:results}

In this appendix, for each problem class we show plots for STT, \tttanneal, and \ttttotal.  Note that the STT plots show results for a single SA solver fixed at 10,000 sweeps.  This is always the best SA solver in terms of STT, though it is only the best SA solver in terms of TTT metrics at the largest problem sizes.  Each plot shows medians for each solver for each problem size with error bars showing 95\% confidence intervals.

The TTT plots in this appendix show the three versions of simulated annealing, which for the sake of cleanliness we combine in the main body.  For RAN1 inputs, \texttt{an\_ms\_r1\_nf} is always the fastest SA implementation.  For other inputs,  our estimated \texttt{an\_ss\_ge\_fi} times are lower by a constant factor than those of our SA code, except on small inputs where our in-house SA is faster by virtue of its lower initialization cost.

\clearpage

\subsection{AC3-odd}

\noindent\includegraphics[width=\textwidth]{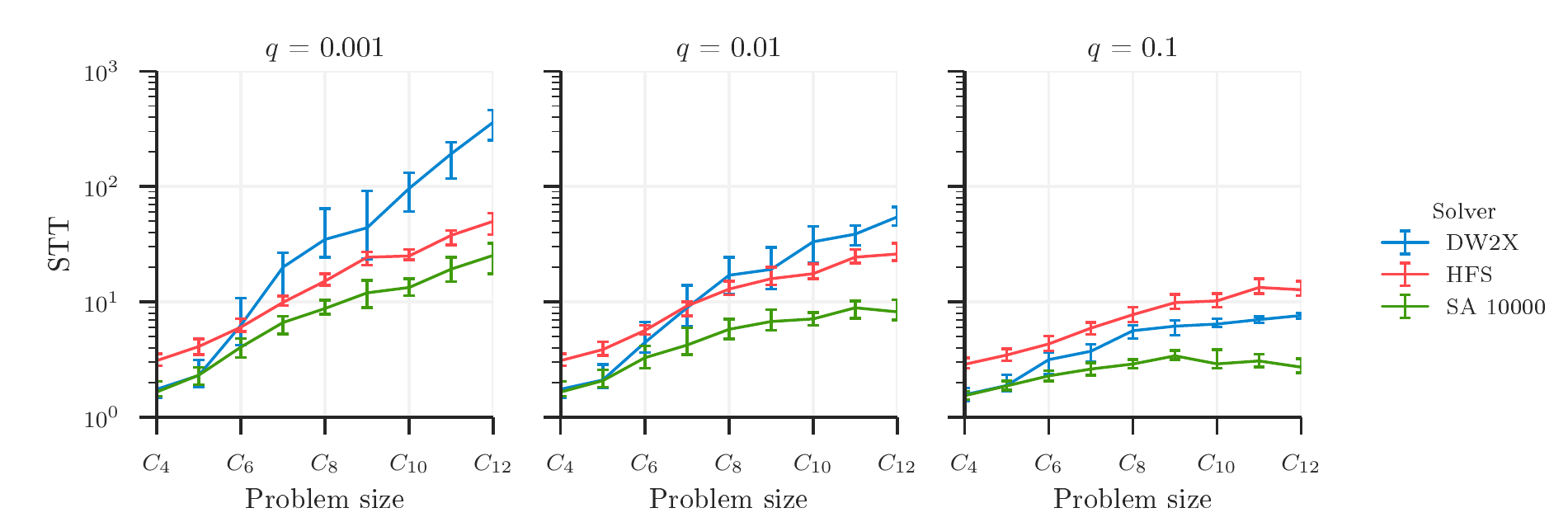}

\hrule\vspace{3mm}

\noindent\includegraphics[width=\textwidth]{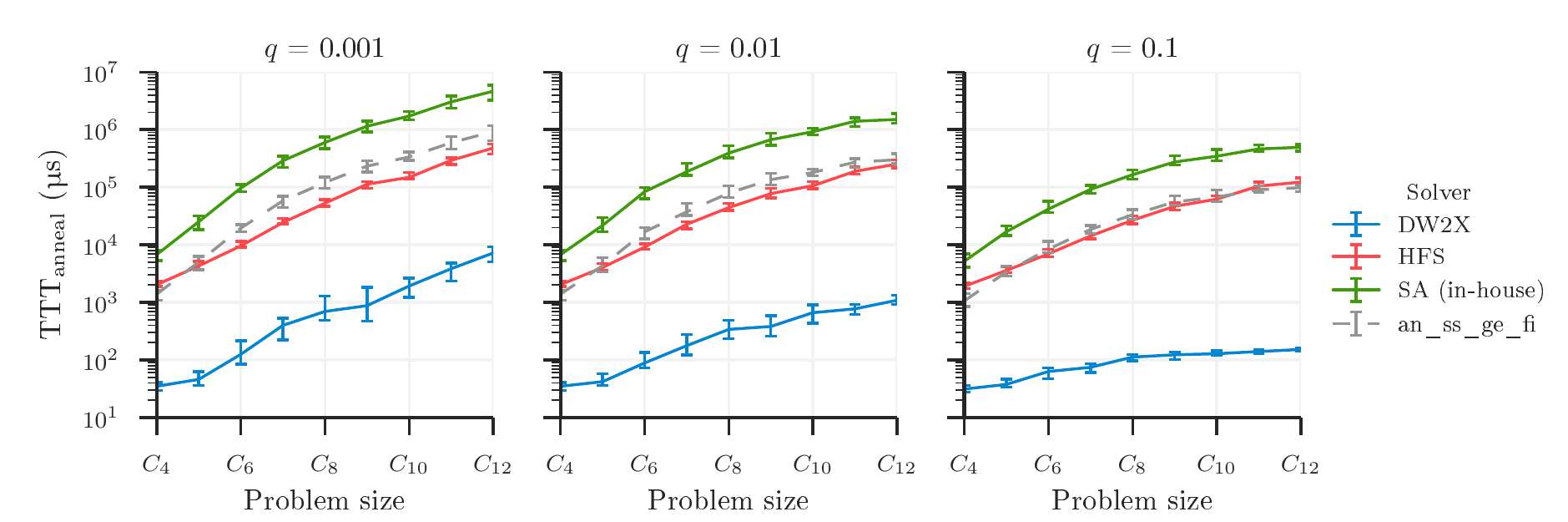}

\hrule\vspace{3mm}

\noindent\includegraphics[width=\textwidth]{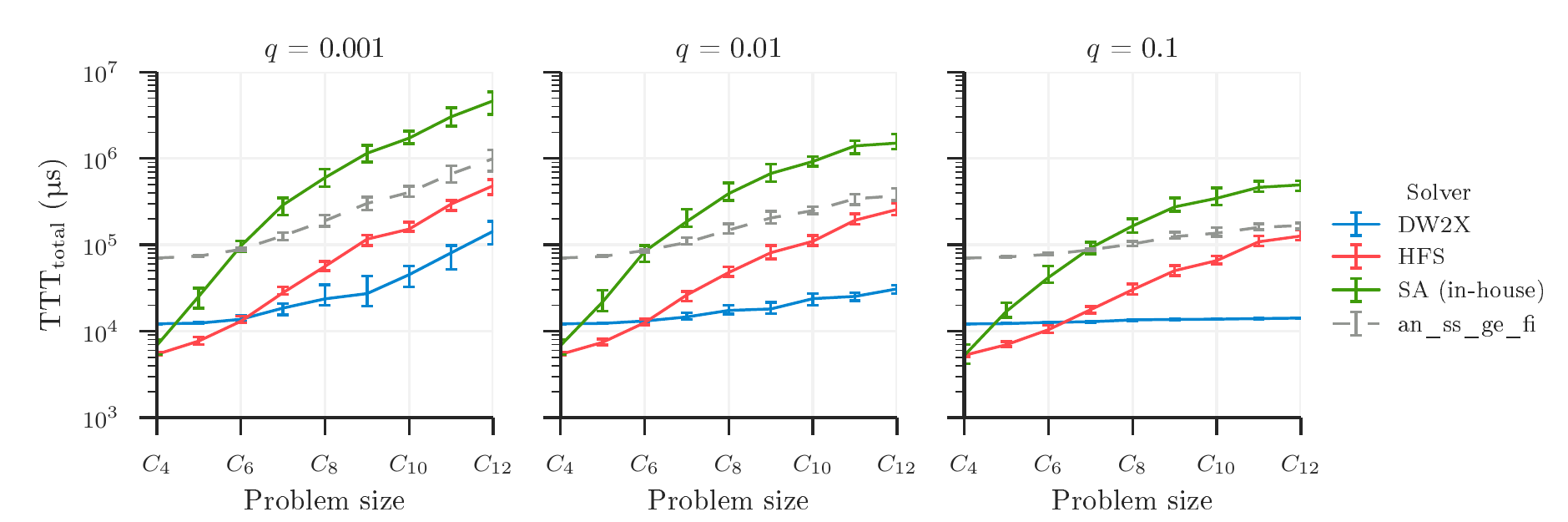}

\subsection{FL3}

\noindent\includegraphics[width=\textwidth]{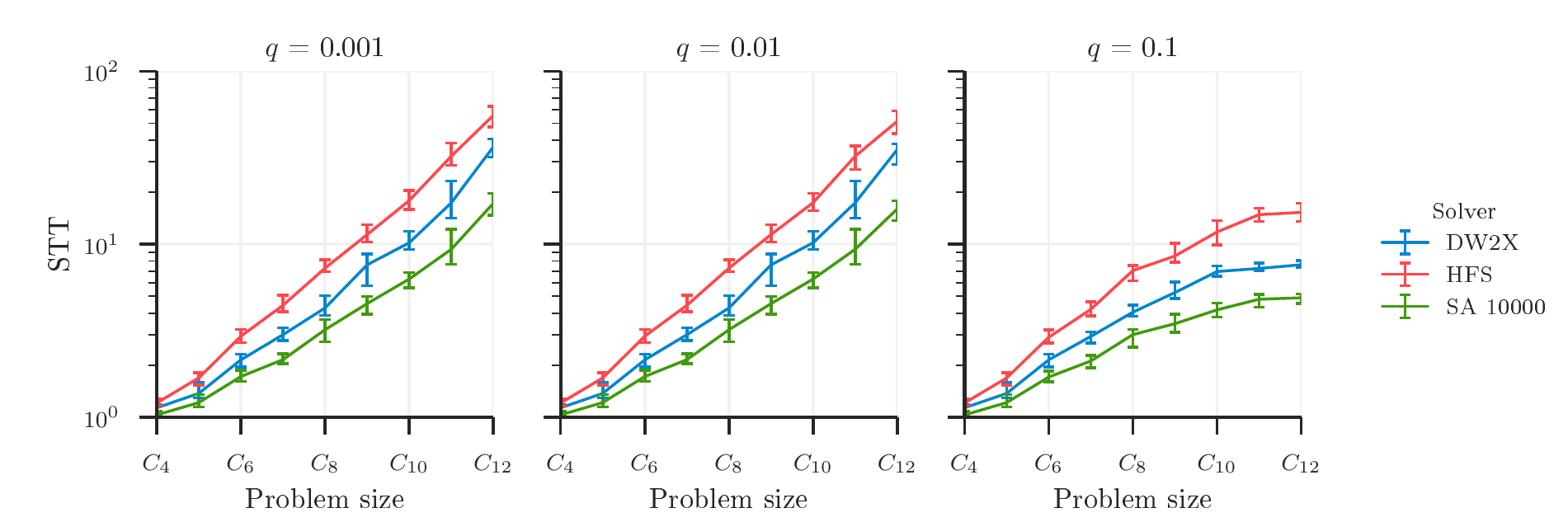}

\hrule\vspace{3mm}

\noindent\includegraphics[width=\textwidth]{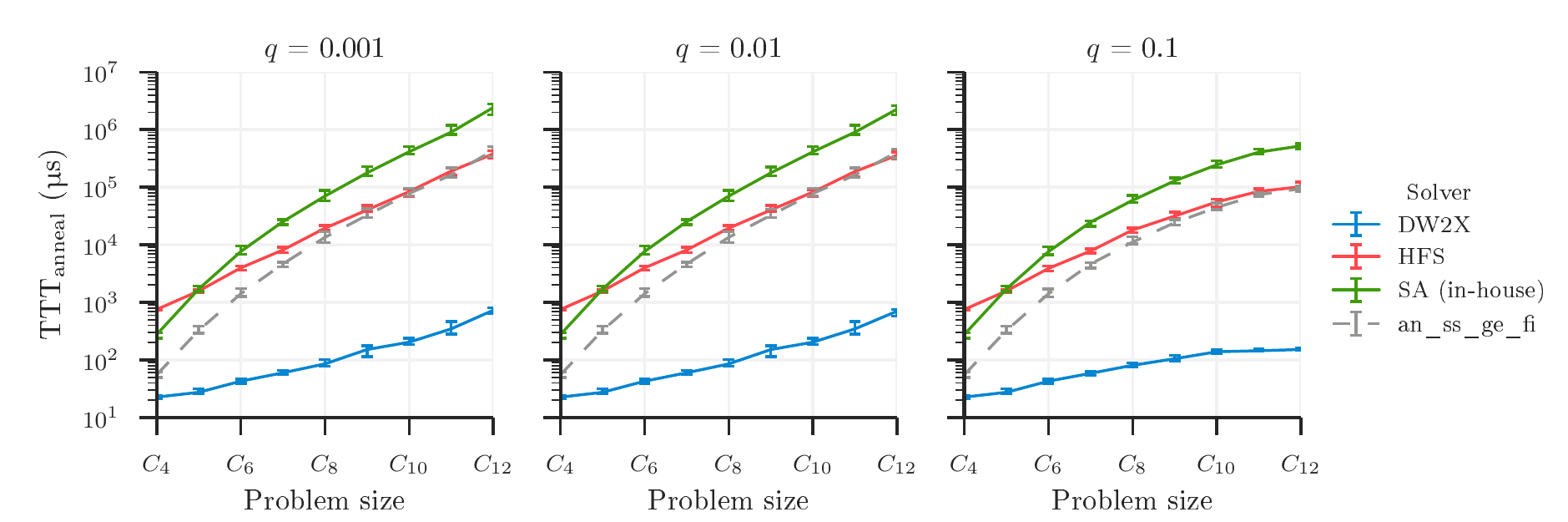}

\hrule\vspace{3mm}

\noindent\includegraphics[width=\textwidth]{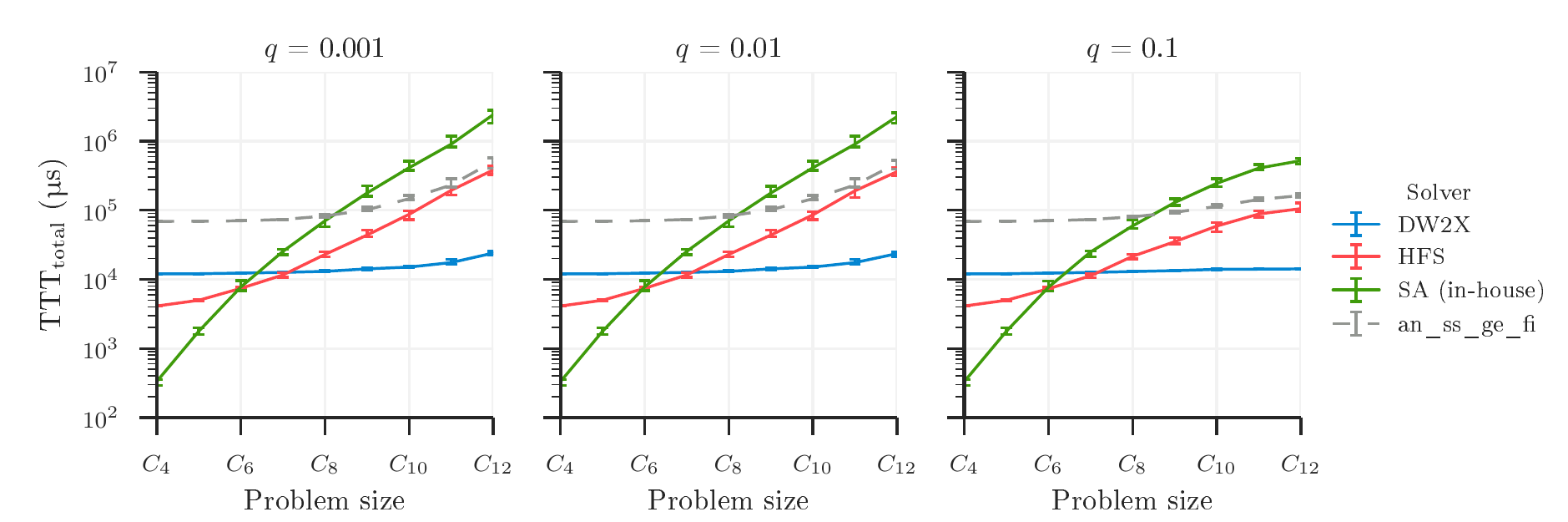}

\subsection{RAN1}

\noindent\includegraphics[width=\textwidth]{figures/RAN1_stt}

\hrule\vspace{3mm}

\noindent\includegraphics[width=\textwidth]{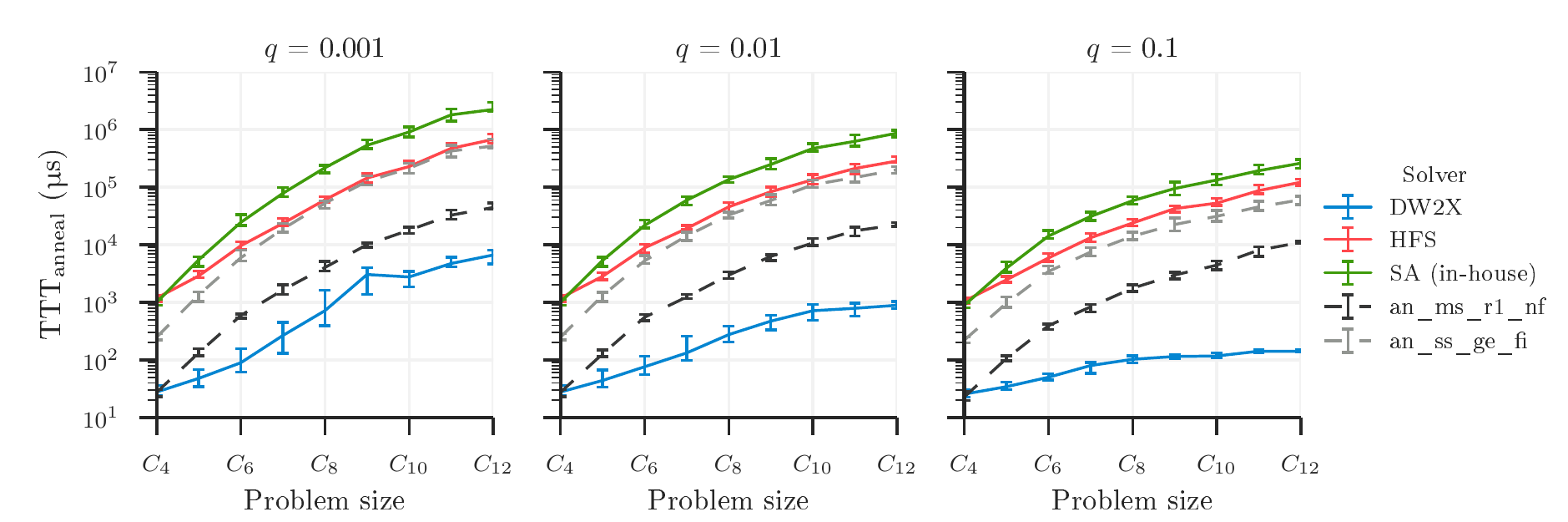}

\hrule\vspace{3mm}

\noindent\includegraphics[width=\textwidth]{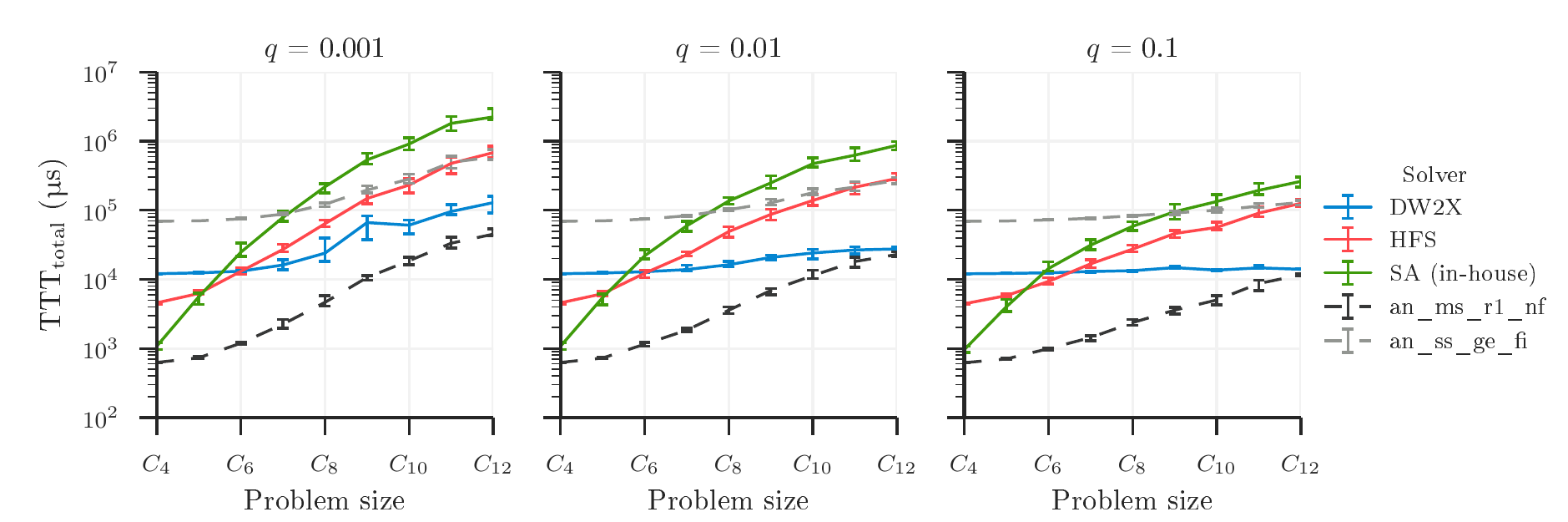}

\subsection{RAN3}

\noindent\includegraphics[width=\textwidth]{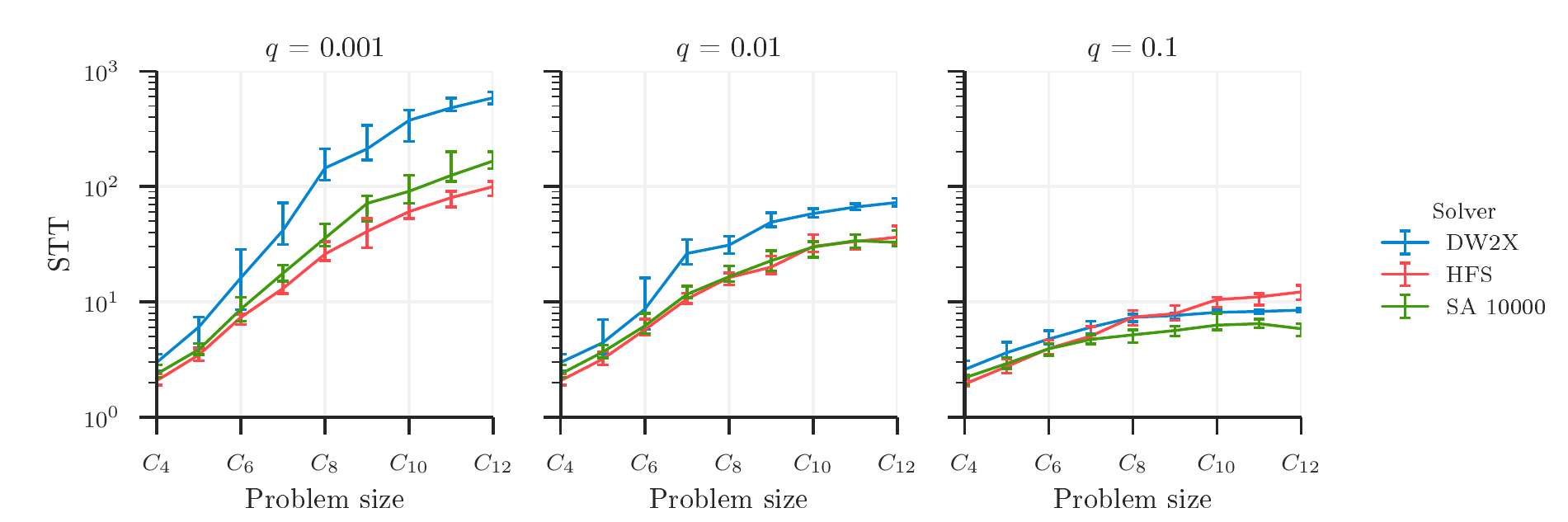}

\hrule\vspace{3mm}

\noindent\includegraphics[width=\textwidth]{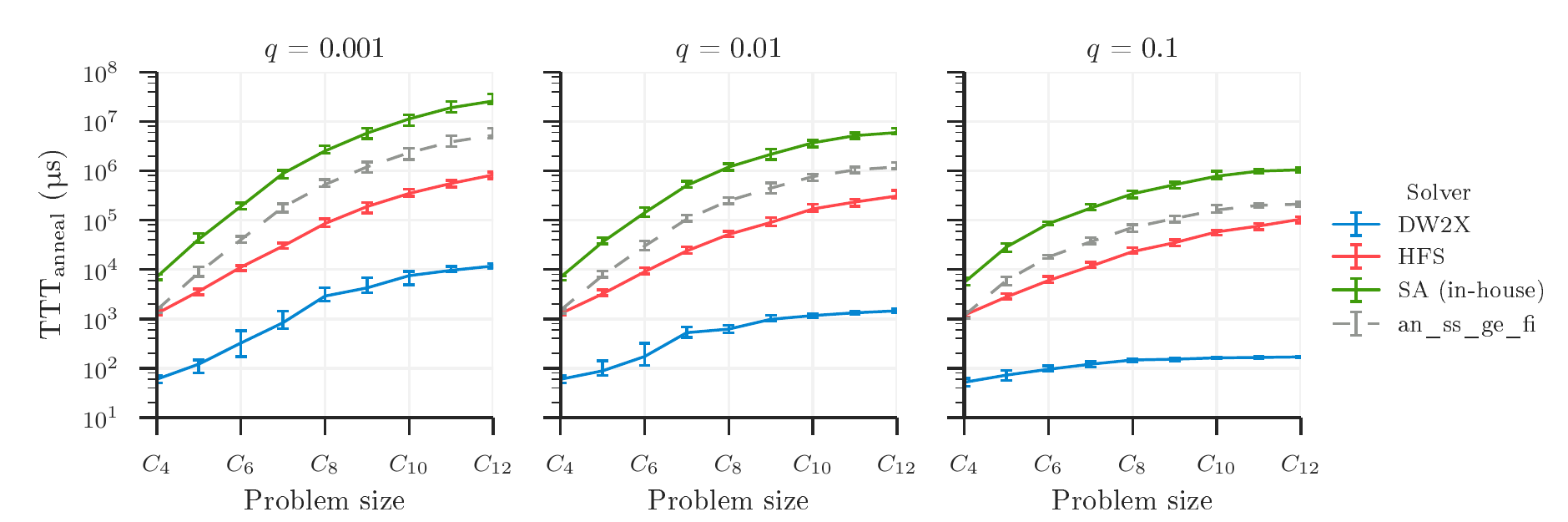}

\hrule\vspace{3mm}

\noindent\includegraphics[width=\textwidth]{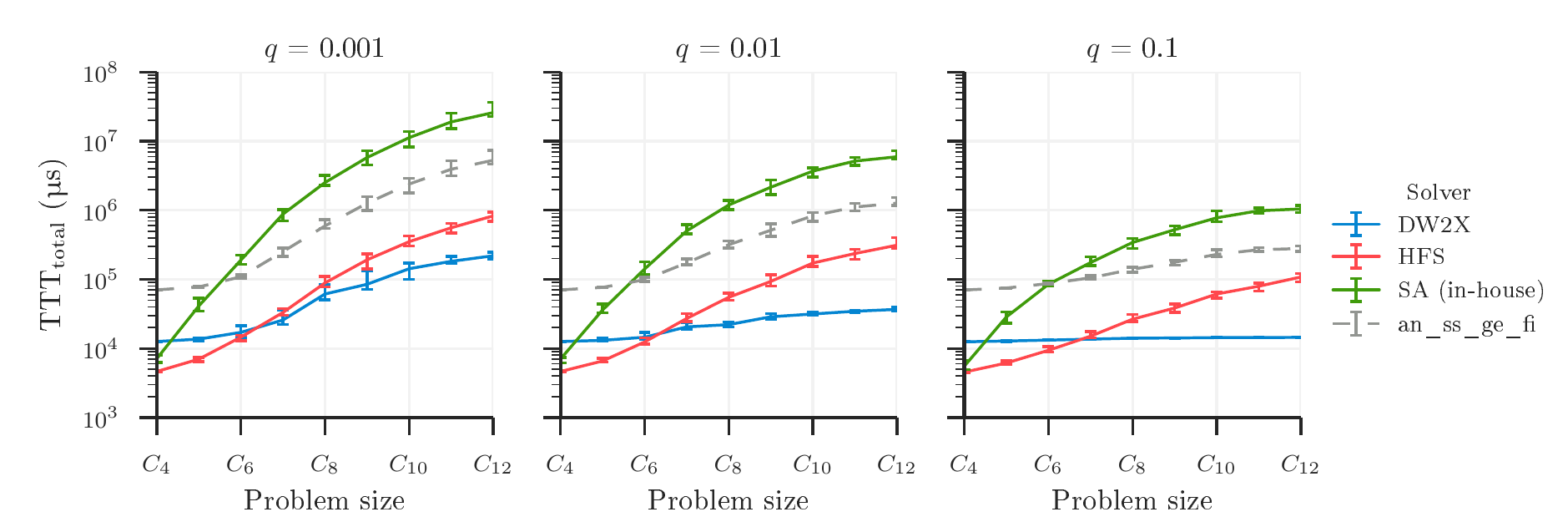}

\subsection{RAN7}

\noindent\includegraphics[width=\textwidth]{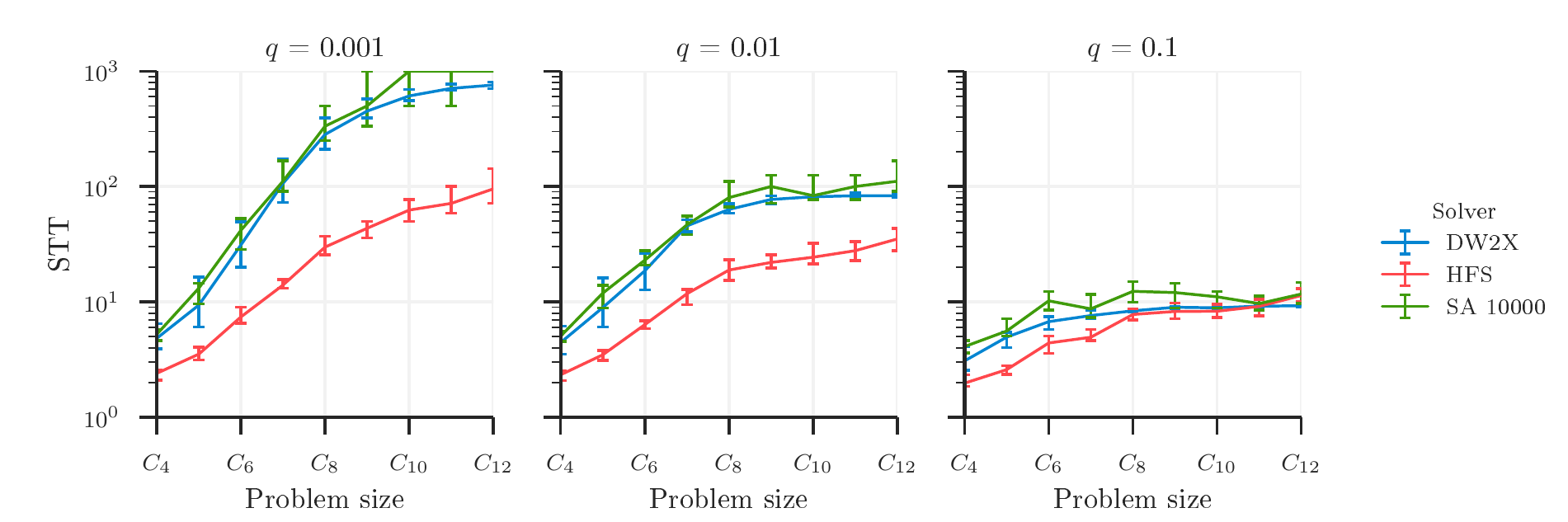}

\hrule\vspace{3mm}

\noindent\includegraphics[width=\textwidth]{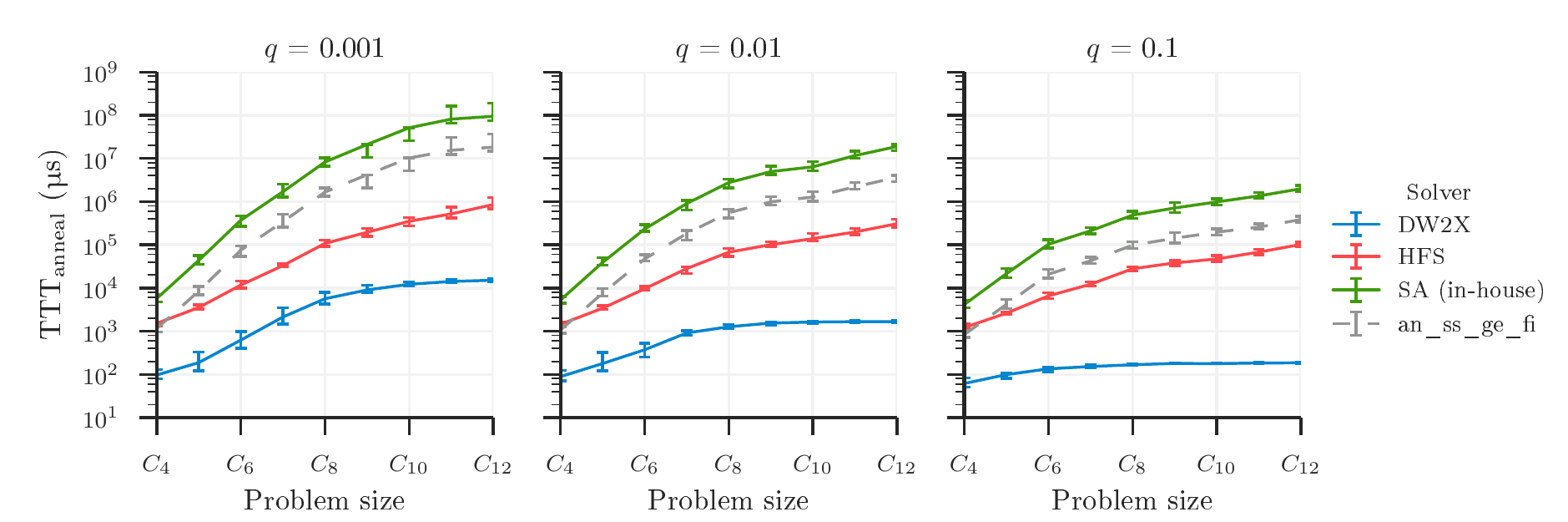}

\hrule\vspace{3mm}

\noindent\includegraphics[width=\textwidth]{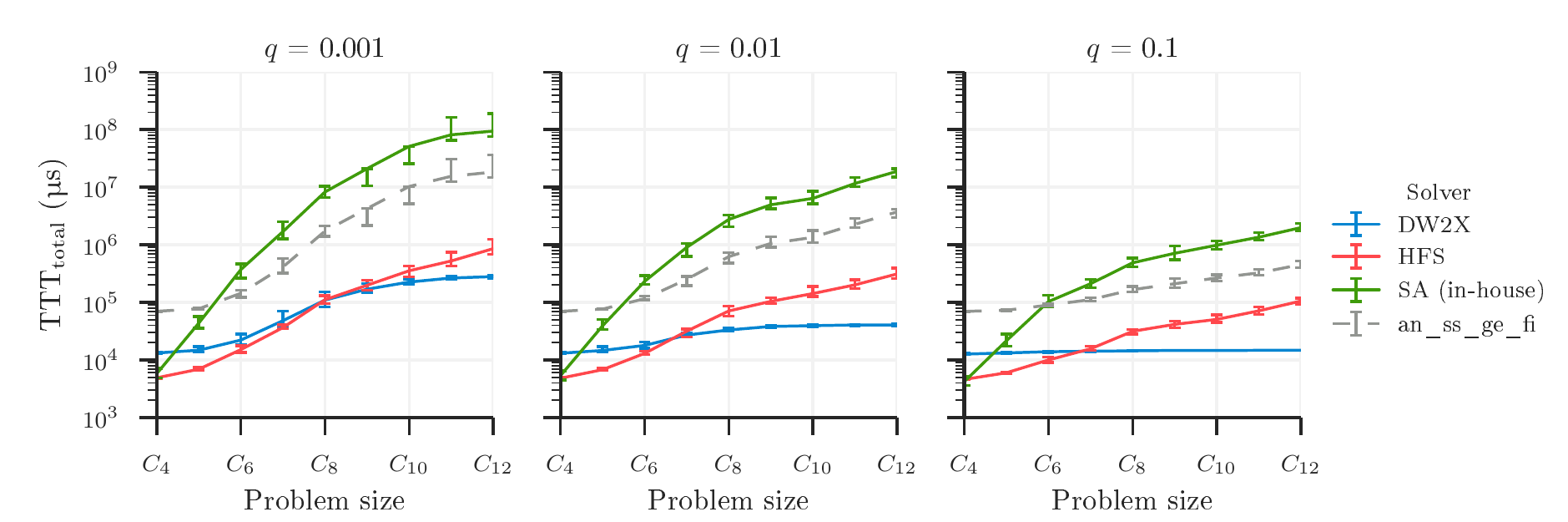}

\subsection{RAN127}

\noindent\includegraphics[width=\textwidth]{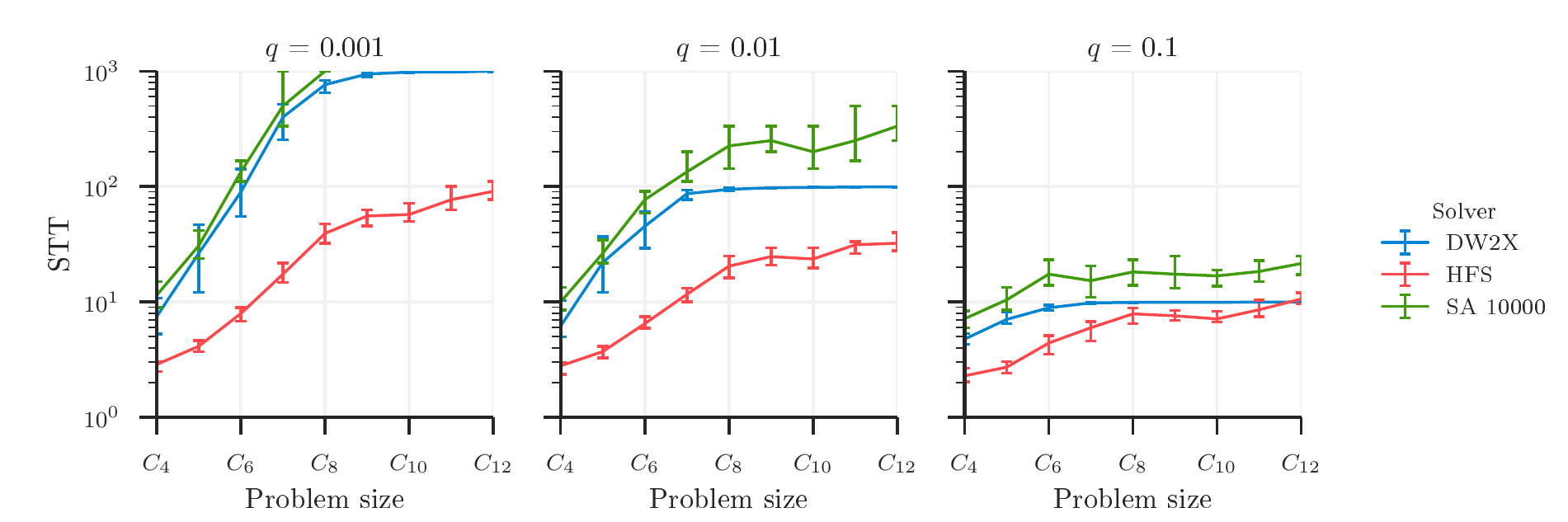}

\hrule\vspace{3mm}

\noindent\includegraphics[width=\textwidth]{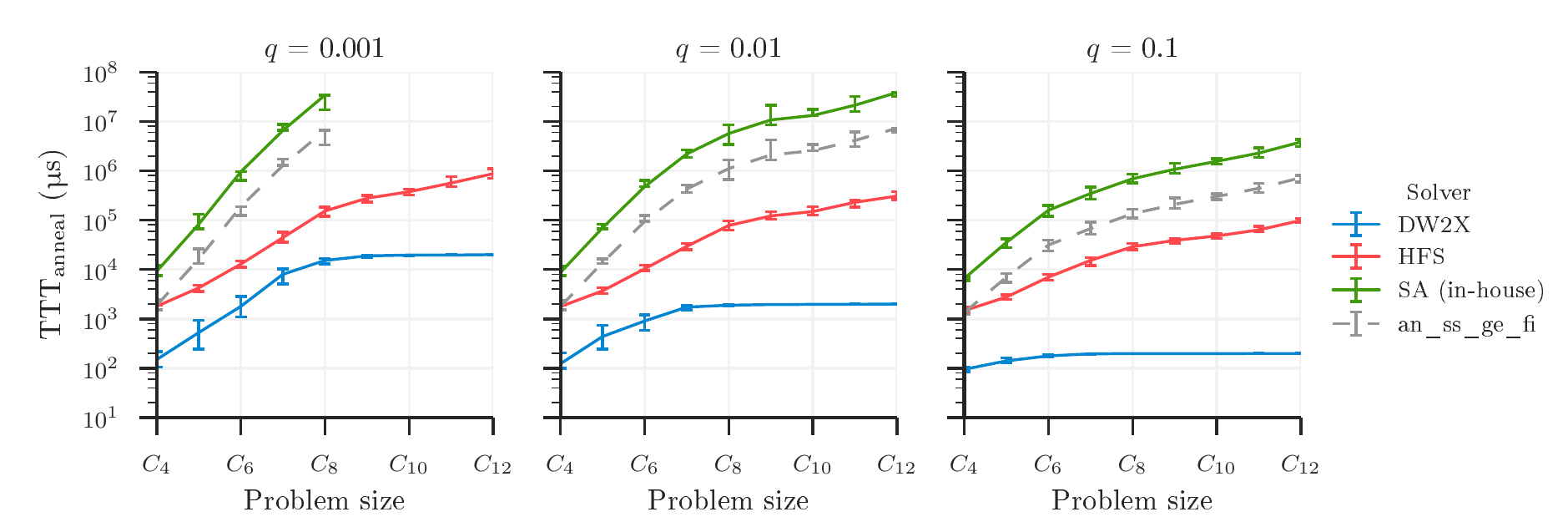}

\hrule\vspace{3mm}

\noindent\includegraphics[width=\textwidth]{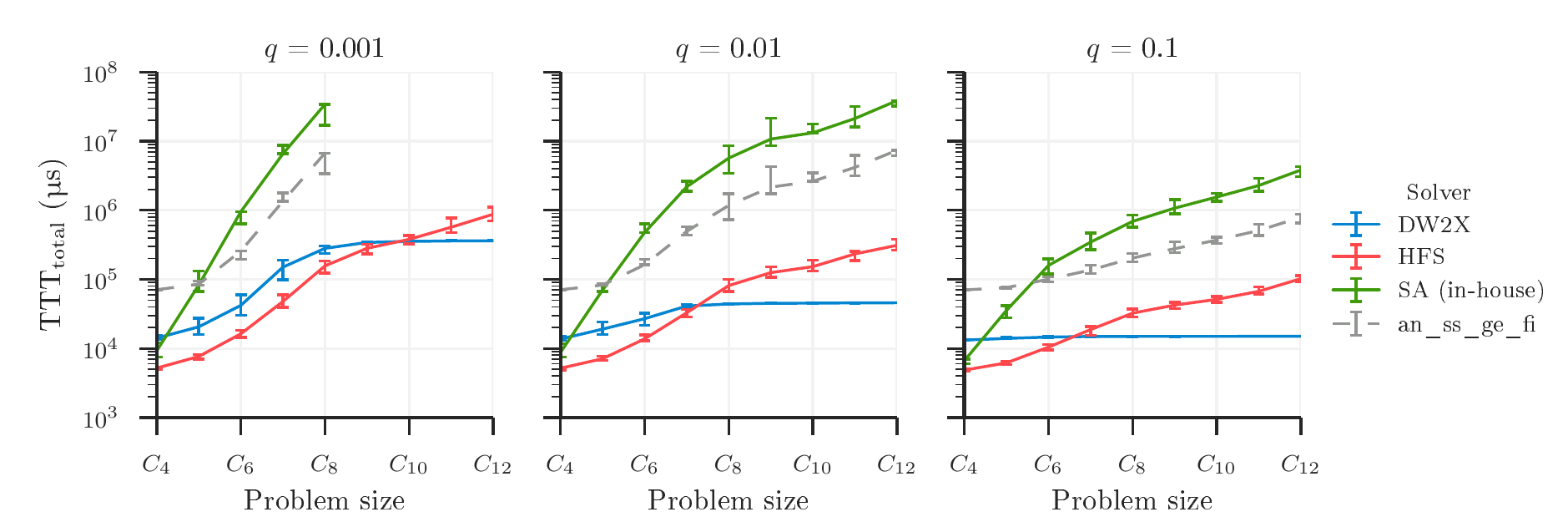}

\end{document}